%% file: main.tex
\documentclass[preprint]{ptephy_v1}

\preprintnumber{OU-HET-1164} 

\usepackage{comment}
\usepackage{physics}
\usepackage{mathrsfs}
\usepackage{braket}
\usepackage[whole]{bxcjkjatype}

\makeatletter
\def\lastpage@putlabel{}
\makeatother
\usepackage{lastpage,hyperref}
\usepackage{breakurl}
\usepackage{subfigure}
\usepackage[utf8]{inputenc}
\usepackage{amsmath, multirow,bigdelim}
\usepackage{mathdots}

\newcommand{\fix}[1]{\textcolor{red}{[#1]}}

\newcommand{\Slash}[1]{{\ooalign{\hfil/\hfil\crcr\(#1\)}}}

\begin{document}

\title{Curved domain-wall fermion and its anomaly inflow}


\author{Shoto Aoki}
\author{Hidenori Fukaya}
\affil{Department of Physics, Osaka University,\\Toyonaka, Osaka 560-0043, Japan \email{saoki@het.phys.sci.osaka-u.ac.jp}\email{hfukaya@het.phys.sci.osaka-u.ac.jp}}






\begin{abstract}
We investigate the effect of $U (1)$ gauge field on lattice fermion systems with a curved domain-wall mass term. In the same way as the conventional flat domain-wall fermion, the chiral edge modes appear localized at the wall, whose Dirac operator contains the induced gravitational potential as well as the $U(1)$ vector potential. In the case of $S^1$ domain-wall fermion on a two-dimensional flat lattice, we find a competition between the Aharonov-Bohm(AB) effect and gravitational gap in the Dirac eigenvalue spectrum, which leads to anomaly of the time-reversal ($T$) symmetry. Our numerical result shows a good consistency with the Atiyah-Patodi-Singer index theorem on a disk inside the $S^1$ domain-wall, which describes the cancellation of the $T$ anomaly between the bulk and edge. When the $U(1)$ flux is squeezed inside one plaquette, and the AB phase takes a quantized value $\pi$ mod $2\pi\mathbb{Z}$, the anomaly inflow drastically changes: the strong flux creates another domain-wall around the flux to make the two zero modes coexist. This phenomenon is also observed in the $S^2$ domain-wall fermion in the presence of a magnetic monopole. We find that the domain-wall creation around the monopole microscopically explains the Witten effect.


\end{abstract}

\subjectindex{xxxx, xxx}

\maketitle

\input{main_introduction}
\input{main_review}

\input{main_S1_weakU1b}

\input{main_WittenEffect}

\input{main_S2_monopole}

\section{Summary}


We have analyzed lattice fermion systems with the spherical $S^1$ and $S^2$ domain-wall mass terms embedded into two- and three-dimensional square lattices, respectively. Putting nontrivial $U(1)$ link variables, we have investigated the Dirac eigenvalue spectrum as well as the profile of the corresponding eigenfunctions to understand the interplay between induced gravity and gauge field on the edge-localized modes.

In sec. \ref{sec:S1_AnomalyInflow}, we have considered the $S^1$ domain-wall fermion on a two-dimensional square lattice. When we put a weak $U(1)$ gauge field inside the domain-wall, the gravitational effect encoded in the gap of the Dirac spectrum competes with the AB effect caused by the $U(1)$ flux. The uniform eigenvalue shift in the Dirac spectrum is consistent with the nonzero $\eta$ invariant of the edge-localized mode, or its time-reversal symmetry anomaly. Then the $\eta$ invariant of the total two-dimensional domain-wall
fermion Dirac operator is consistent with the APS index theorem in continuum theory on the two-dimensional disk, with which the $T$ symmetry of the whole system is protected.

When we shrink the $U(1)$ flux to form a vortex, the AB effect on the edge-localized modes is unchanged but the intense gauge field inside makes a drastic change in the bulk. We have revealed an exponentially localized mode at the flux, which has an opposite chirality to the edge modes on the $S^1$ domain-wall. In our microscopical analysis, this mode can be interpreted as an edge mode sitting at another small domain-wall dynamically created by the additive mass renormalization via the Wilson term around the vortex.

In particular, the two Dirac eigenvalues, one is located at the $S^1$ domain-wall, and the other localized at the new domain-wall around the vortex, approach to zero when the AB phase is $\pi$, which is the $T$-symmetric point. Pairing of the two zero modes reflects the fact that the mod two index is a cobordism invariant: they must appear in pairs, to make the total index trivial, on a manifold that is a boundary of a higher dimensional manifold. When we regard the Dirac operator as a Hamiltonian in $d=2$ space dimensions, rather than in $1+1$ space-time dimensions, the appearance of the two zero modes and their small split due to tunneling between two domain-walls gives a microscopic description how the vortex gains a fractional electric charge.

We have also analyzed the $S^2$ domain-wall fermion system in the presence of a magnetic monopole at the origin. The Atiyah-Singer index theorem on the two-dimensional sphere indicates a zero edge localized mode on it. The axial $U(1)$ anomaly cancellation of the total three-dimensional system is then required as the same cobordism argument as the $S^1$ domain-wall, and suggests a dynamical creation of another domain-wall around the monopole.

We have numerically confirmed the sign flip of the effective fermion mass
around the monopole, and thus creation of the domain-wall, on which
the edge-localized zero-mode appears.
We have also found that 50\% of the zero mode's amplitude is localized at
the monopole, while the other 50\% is at the $S^2$ domain-wall.
We believe that the analysis gives a microscopic description of the
Witten effect that the monopole gains a 1/2 charge of an electron.
We are now trying to analytically confirm the results exactly solving
the ``negatively" massive Dirac equation with a Wilson term \cite{AFKKMinpre}.

\section*{Acknowledgment}

We thank M. Furuta, K. Hashimoto, S. Iso, M. Kawahira, N. Kan, M. Koshino, Y. Matsuki, S. Matsuo, T. Onogi, S. Yamaguchi, M. Yamashita and R. Yokokura for useful discussions. 
This work was supported by JST SPRING, Grant Number JPMJSP2138, and in part by JSPS KAKENHI Grant Number JP18H01216, JP18H04484 and JP22H01219.

\bibliographystyle{JHEP}
\bibliography{ref}

\appendix

\section{Edge mode at $S^1$ domain-wall with the $U(1)$ flux}

\label{sec:Eigenvalue of Edge mode S1}

In this section, we analytically solve the eigenvalue of edge modes of
\begin{align}
    H =\sigma_3 \qty(\sum_{i=1,2}\sigma_j (\partial_j -iA_j)  +M), 
\end{align}
where $M=m\text{sign}(r-r_0)$ and $A$ is given by \eqref{eq:U1 flux on S1}. This operator is the continuum limit of \eqref{eq:Hermitian Wilson Dirac op of S1 in R2} and acts on a periodic spinor field. Setting $A=\alpha d\theta$ on the whole of the two-dimensional Euclidean space, the operator becomes 
\begin{align}
    H=\sigma_3 \mqty( M & e^{-i\theta} \qty( \pdv{}{r}-i \frac{1}{r} \pdv{}{\theta} -\frac{\alpha}{r} ) \\ e^{i\theta} \qty( \pdv{}{r}+i \frac{1}{r} \pdv{}{\theta} +\frac{\alpha}{r} ) & M ).
\end{align}

This operator commutes with the total angular momentum $J=-i\pdv{}{\theta}+ \frac{1}{2} \sigma_3$, whose eigenvalue takes $j\in \frac{1}{2}+\mathbb{Z}$. The eigenfunction with $J=j$ localized at the wall is given by
\begin{align}
    \psi^{E,j}&=\left\{
\begin{array}{ll}
A \mqty(\sqrt{m^2-E^2} I_{ \abs{j-\frac{1}{2}-\alpha}} (\sqrt{m^2-E^2} r)e^{i(j-\frac{1}{2})\theta }\\ 
         (m+E) I_{\abs{j+\frac{1}{2}-\alpha}} (\sqrt{m^2-E^2} r)e^{i(j+\frac{1}{2})\theta }) & (r<r_0) \\
B\mqty((m+E)K_{j-\frac{1}{2}-\alpha} (\sqrt{m^2-E^2} r)e^{i(j-\frac{1}{2})\theta }\\ 
         \sqrt{m^2-E^2} K_{j+\frac{1}{2}-\alpha} (\sqrt{m^2-E^2} r)e^{i(j+\frac{1}{2})\theta }) & (r>r_0).
\end{array}
\right.
\end{align}
The connection condition $r=r_0$
\begin{align}\label{eq:condition of E}
    \frac{I_{\abs{j-\frac{1}{2}-\alpha}}}{I_{\abs{j+\frac{1}{2}-\alpha}}}\frac{K_{j+\frac{1}{2}-\alpha}}{K_{j-\frac{1}{2}-\alpha}}(\sqrt{m^2-E^2}r_0)=\frac{m+E}{m-E}, 
\end{align}
determines the eigenvalue $E$. In the large mass limit or $m\gg E$, the eigenvalue converges to
\begin{align}
    E\simeq \frac{j-\alpha}{r_0}\ \qty(j=\pm\frac{1}{2},\pm\frac{3}{2},\cdots) \label{eq:S1 eigenavlue}. 
\end{align}

\section{Eigenstate localized at the $U(1)$ flux}
\label{sec:Eigenvalue of localized mode S1}

In this appendix, we analytically show that a domain-wall is dynamically created near the vortex, which captures a bound state of an electron. Here, we ignore the outside of the $S^1$ domain-wall or take the $r_0 \to \infty$ limit and consider the operator 
\begin{align} \label{eq:approximation of H S1}
    H=\sigma_3 \qty( \sigma_j D_j -\frac{1}{2M_{PV} } D^2+ M), 
\end{align}
where $M=-m<0$, $M_{PV}=1/a$ and $D_j= \partial_j-iA_j$ is a covariant derivative in $j$-th direction. Note that the Wilson term is put by hand, in order to make the sign of fermion mass well-defined \cite{Zhao2012Amagneticmonopole}. This operator also commutes with $J=-i\pdv{}{\theta}+ \frac{1}{2} \sigma_3$. The eigenfunction can be written as $\psi= \mqty(f(r) e^{i(j-\frac{1}{2})\theta} \\ g(r) e^{i(j+\frac{1}{2})\theta})$.

\subsection{Exterior $(r>r_1)$}

In the region of $r>r_1$, the operator \eqref{eq:approximation of H S1} is given by
\begin{align}
    H=\sigma_3 \mqty( M-\frac{D^2}{2M_{PV}} & e^{-i\theta} \qty( \pdv{}{r}-i \frac{1}{r} \pdv{}{\theta} -\frac{\alpha}{r} ) \\ e^{i\theta} \qty( \pdv{}{r}+i \frac{1}{r} \pdv{}{\theta} +\frac{\alpha}{r} ) & M -\frac{D^2}{2M_{PV}} ),
\end{align}
where $D^2= \pdv[2]{}{r}+\frac{1}{r}\pdv{}{r}- \frac{1}{r^2} \qty( i\pdv{}{\theta} +\alpha)^2$. Then the equation in the $r$-direction is
\begin{align}
    E\mqty( f \\ g) = \mqty( M + \frac{a_j^\dagger a_j}{2M_{PV}} & a_j^\dagger \\ a_j & -M - \frac{a_j a_j^\dagger }{2M_{PV}} )  \mqty( f \\ g),
\end{align}
where $E$ is an eigenvalue of the localized mode. $a_j$ and $a_j^\dagger$ are a differential operator defined by
\begin{align}
    a_{j}=-\pdv{}{r} +\frac{j-\frac{1}{2} -\alpha}{r},~a_{j}^\dagger= \pdv{}{r} +\frac{j+\frac{1}{2} -\alpha}{r}.
\end{align}
Assuming $a_j f\propto g $ and $ a_j^\dagger g \propto f$, we find a set of solutions in the term $f=a K_{j-\frac{1}{2}-\alpha}(\kappa r)$ and $g=b K_{j+\frac{1}{2}-\alpha} (\kappa r)$ for a complex number $a$, $b$ and $\kappa$. The coefficients $a$ and $b$ satisfy 
\begin{align}
E\mqty(a\\ b)=  \mqty(  M -\frac{\kappa^2}{2M_{PV}} & -\kappa \\ +\kappa & -\qty(M -\frac{\kappa^2}{2M_{PV}})) \mqty(a\\ b).
\end{align}
There are two $\kappa$'s that satisfy the above equation:
\begin{align}
    \frac{\kappa^2}{M_{PV}^2}=\frac{\kappa_\pm^2}{M_{PV}^2}=\qty( \pm 1 +\sqrt{1+2 \frac{M}{M_{PV}} +\frac{E^2}{M_{PV}^2}})^2-\frac{E^2}{M_{PV}^2}.
\end{align}
Thus two eigenstates are
\begin{align}
    \psi_{o\pm}=\mqty(\qty(M -\frac{\kappa_{\pm}^2}{2M_{PV}} +E) K_{j-\frac{1}{2}-\alpha}(\kappa_{\pm} r) e^{i(j-\frac{1}{2})\theta} \\ \kappa_{\pm} K_{j+\frac{1}{2}-\alpha}(\kappa_{\pm} r) e^{i(j+\frac{1}{2})\theta} ).
\end{align}

\subsection{Interior $(r<r_1)$}
For $r<r_1$ the Dirac operator is given by
\begin{align}
    H=\sigma_3 \mqty( M-\frac{D^2}{2M_{PV}} & e^{-i\theta} \qty( \pdv{}{r}-i \frac{1}{r} \pdv{}{\theta} -\frac{\alpha r}{r_1^2} ) \\ e^{i\theta} \qty( \pdv{}{r}+i \frac{1}{r} \pdv{}{\theta} +\frac{\alpha r}{r_1^2} ) & M -\frac{D^2}{2M_{PV}} ),
\end{align}
where $D^2=\pdv[2]{}{r}+\frac{1}{r}\pdv{}{r} -\frac{1}{r^2} \qty(i\pdv{}{\theta}+ \alpha \frac{r^2}{r_1^2} )^2$. $f$ and $g$ satisfy
\begin{align}
    E\mqty( f \\ g) = \mqty( M + \frac{1}{2M_{PV}}\qty( a_j^\dagger a_j+ 2\frac{\alpha}{r_1^2}) & a_j^\dagger \\ a_j & -M - \frac{1}{2M_{PV}}\qty(a_j a_j^\dagger - 2\frac{\alpha}{r_1^2}) )  \mqty( f \\ g),
\end{align}
where $a_j$ and $a_j^\dagger$ are defined by
\begin{align}
    a_{j}=-\pdv{}{r} +\frac{1 }{r} \qty(j-\frac{1}{2} - \alpha \frac{r^2}{r_1^2}) ,~a_{j}^\dagger= \pdv{}{r} + \frac{1 }{r}\qty(j+\frac{1}{2} - \alpha \frac{r^2}{r_1^2}).
\end{align}
Assuming that $f=aF$ is a eigenfunction of $a_j^\dagger a_j F =LF$. We find
\begin{align}
    F=\left\{ \begin{array}{ll}
        r^{j-\frac{1}{2}} e^{- \frac{ \alpha r^2}{ 2r_1^2}} {}_1 F_1( - \frac{r_1^2}{4 \alpha} L, j+ \frac{1}{2}; \alpha \frac{r^2}{r_1^2}) & (j=\frac{1}{2}, \frac{3}{2},\cdots) \\
       r^{-j+\frac{1}{2}} e^{- \frac{ \alpha r^2}{ 2r_1^2}} {}_1 F_1(-j+\frac{1}{2} - \frac{r_1^2}{4 \alpha} L, -j+ \frac{3}{2}; \alpha \frac{r^2}{r_1^2})  & (j=-\frac{1}{2}, -\frac{3}{2},\cdots)
    \end{array} \right. 
\end{align}
an overall constant. ${}_1 F_1(a,b;z)=\sum_{n=0}^\infty \frac{\Gamma( a+n)}{ \Gamma( a)} \frac{\Gamma( b)}{\Gamma( b+n)} \frac{z^n}{ n!}$ is a confluent hypergeometric function which satisfies
\begin{align}
    \qty(z \dv[2]{}{z} + (b-z) \dv{}{z} -a) {}_1 F_1(a,b;z) =0.
\end{align}

Denoting $g$ by $b a_j F$, we obtain an equation for $a$ and $b$,
\begin{align}
    E\mqty( a \\ b) = \mqty( M + \frac{1}{2M_{PV}}\qty( L+ 2\frac{\alpha}{r_1^2}) & L\\ 1 & -M - \frac{1}{2M_{PV}}\qty(L - 2\frac{\alpha}{r_1^2}) )  \mqty( a \\ b).
\end{align}
Similarly to the case $r>r_1$, we have two solution for $L$:
\begin{align}
    \frac{L}{M_{PV}^2}= \frac{L_{\pm}}{M_{PV}^2}=-\qty( \pm 1 +\sqrt{1+2 \frac{M}{M_{PV}} +\frac{\qty(E- \frac{\alpha}{ M_{PV} r_1^2} )^2}{M_{PV}^2}})^2+\frac{\qty(E- \frac{\alpha}{ M_{PV} r_1^2} )^2}{M_{PV}^2}
\end{align}
and
\begin{align}
    \psi_{i\pm}=\mqty(\qty(M +\frac{L_\pm}{2M_{PV}} +E- \frac{\alpha}{ M_{PV} r_1^2}) F(r) e^{i(j-\frac{1}{2})\theta} \\ a_j F(r) e^{i(j+\frac{1}{2})\theta} ).
\end{align}

\subsection{Continuity at $r=r_1$}

The total eigenfunction $\psi$ is a liner combination of $\psi_{i\pm}$ when $r<r_1$ and $\psi_{o\pm}$ when $r>r_1$:
\begin{align}
    \psi= \left\{ \begin{array}{cc}
        c_1 \psi_{i+}+ c_2 \psi_{i-}  & (r<r_1) \\
        d_1 \psi_{o+}+ d_2 \psi_{o-} &  (r>r_1)
    \end{array} \right. .
\end{align}
Since $\psi$ and $\pdv{}{r} \psi$ are a continuous at $r=r_1$ and $\psi$ is normalized, we can determine $c_1,~c_2,~d_1,~d_2$ and the eigenvalue $E$. 

The connection of the outer/inner solutions requires $\psi$ and its derivative $\partial_r\psi$ are continuous at $r=r_1$. Knowing the fact that the Hamiltonian relates the eigenfunction $\psi$ and its derivative $\partial_r\psi$ to the operation of the Laplacian $D_\mu D^\mu \psi$, we replace the continuity condition for $\partial_r\psi$ by that for $D_\mu D^\mu \psi$.



We furthermore assume $1/r_1 \gg M_{PV} \gg \abs{M}$ and $j>0$, then the functions are approximated as
\begin{align}
    \psi_{i+}& \simeq \mqty( -\frac{2\alpha}{ M_{PV} r_1^2 }   {}_1 F_1(+\frac{1}{2}, j+ \frac{1}{2}; \alpha \frac{r^2}{r_1^2}) e^{i(j-\frac{1}{2})\theta} \\ -\frac{\alpha}{ r_1 (j+1/2) } {}_1 F_1(+\frac{3}{2}, j+ \frac{1}{2}; \alpha \frac{r^2}{r_1^2}) e^{i(j+\frac{1}{2})\theta} ) r^{ j -\frac{1}{2}}e^{- \frac{ \alpha r^2}{ 2r_1^2}}, \\
    D^2 \psi_{i+} & \simeq \mqty( \frac{2\alpha}{ M_{PV} r_1^2 } M_{PV}^2  {}_1 F_1(+\frac{1}{2}, j+ \frac{1}{2}; \alpha \frac{r^2}{r_1^2}) e^{i(j-\frac{1}{2})\theta} \\ -\frac{\alpha}{ r_1 (j+1/2) } \frac{4\alpha}{r_1^2} {}_1 F_1(+\frac{3}{2}, j+ \frac{1}{2}; \alpha \frac{r^2}{r_1^2}) e^{i(j+\frac{1}{2})\theta} ) r^{ j -\frac{1}{2}} e^{- \frac{ \alpha r^2}{ 2r_1^2}},
\end{align}

\begin{align}
    \psi_{i-}& \simeq \mqty(-\frac{M_{PV}}{2}   {}_1 F_1(-\frac{1}{2}, j+ \frac{1}{2}; \alpha \frac{r^2}{r_1^2}) e^{i(j-\frac{1}{2})\theta} \\ \frac{\alpha}{ r_1 (j+1/2) } {}_1 F_1(+\frac{1}{2}, j+ \frac{1}{2}; \alpha \frac{r^2}{r_1^2}) e^{i(j+\frac{1}{2})\theta} )r^{ j -\frac{1}{2}} e^{- \frac{ \alpha r^2}{ 2r_1^2}}, \\
    D^2 \psi_{i-} & \simeq \mqty( -\frac{M_{PV}}{2} \frac{4\alpha}{r_1^2}  {}_1 F_1(-\frac{1}{2}, j+ \frac{1}{2}; \alpha \frac{r^2}{r_1^2}) e^{i(j-\frac{1}{2})\theta} \\ -\frac{\alpha}{ r_1 (j+1/2) } M_{PV}^2 {}_1 F_1(+\frac{1}{2}, j+ \frac{1}{2}; \alpha \frac{r^2}{r_1^2}) e^{i(j+\frac{1}{2})\theta} ) r^{ j -\frac{1}{2}} e^{- \frac{ \alpha r^2}{ 2r_1^2}},
\end{align}

\begin{align}
\psi_{o+} &\simeq \mqty( -2M_{PV} K_{j-\frac{1}{2}-\alpha}(2M_{PV} r) e^{i(j-\frac{1}{2})\theta} \\ 2M_{PV} K_{j+\frac{1}{2}-\alpha}(2 M_{PV} r) e^{i(j+\frac{1}{2})\theta} ),\\
D^2\psi_{o+} &\simeq \mqty( (2M_{PV})^3 K_{j-\frac{1}{2}-\alpha}(2M_{PV} r) e^{i(j-\frac{1}{2})\theta} \\ -(2M_{PV})^3 K_{j+\frac{1}{2}-\alpha}(2 M_{PV} r) e^{i(j+\frac{1}{2})\theta} ),
\end{align}

\begin{align}
\psi_{o-} &\simeq \mqty( (M+E) K_{j-\frac{1}{2}-\alpha}(\sqrt{M^2 -E^2} r) e^{i(j-\frac{1}{2})\theta} \\ \sqrt{M^2 -E^2} K_{j+\frac{1}{2}-\alpha}(\sqrt{M^2 -E^2} r) e^{i(j+\frac{1}{2})\theta} ),\\
D^2\psi_{o-} &\simeq \mqty( -(M+E) \sqrt{M^2 -E^2}^2 K_{j-\frac{1}{2}-\alpha}(\sqrt{M^2 -E^2} r) e^{i(j-\frac{1}{2})\theta} \\ -\sqrt{M^2 -E^2}^3 K_{j+\frac{1}{2}-\alpha}(\sqrt{M^2 -E^2} r) e^{i(j+\frac{1}{2})\theta} ).
\end{align}

From the continuity, $E$ must satisfy
\small
\begin{align}
    \det \mqty( -\frac{2\alpha}{ M_{PV} r_1^2 } u_{\frac{1}{2}}  & -\frac{M_{PV}}{2} v_{-\frac{1}{2}}   & -2M_{PV} k_0( 2M_{PV} r_1)   & (M+E) k_0( \sqrt{M^2 -E^2}  r_1) \\
-\frac{\alpha}{ r_1 (j+1/2) } u_{\frac{3}{2}}  & \frac{\alpha}{ r_1 (j+1/2) }  v_{\frac{1}{2}} & 2M_{PV}  k_1(2M_{PV} r_1) & \sqrt{M^2 -E^2} k_1 (\sqrt{M^2 -E^2} r_1) \\
\frac{2\alpha M_{PV}}{ r_1^2 }  u_\frac{1}{2} & -\frac{M_{PV}}{2} \frac{4\alpha}{r_1^2} v_{-\frac{1}{2}} & (2M_{PV})^3  k_0(2M_{PV} r_1) & -(M+E) \sqrt{M^2 -E^2}^2 k_0(\sqrt{M^2 -E^2} r_1) \\
-\frac{4\alpha^2}{ r_1^3 (j+1/2) } u_\frac{3}{2} & -\frac{\alpha M_{PV}^2}{ r_1 (j+1/2) } v_\frac{1}{2} & -(2M_{PV})^3 k_1(2M_{PV} r_1)  &  -\sqrt{M^2 -E^2}^3 k_1(\sqrt{M^2 -E^2} r_1)  )=0,
\end{align}
\normalsize
where $u_{\frac{1}{2}+n}= {}_1 F_1(+\frac{1}{2}+n, j+ \frac{1}{2}+n; \alpha )e^{- \frac{ \alpha }{ 2}} r_1^{ j -\frac{1}{2}} \simeq r_1^{ j -\frac{1}{2}},~v_{-\frac{1}{2}+n} = {}_1 F_1(-\frac{1}{2}+n, j+ \frac{1}{2}+n; \alpha )e^{- \frac{ \alpha }{ 2}} r_1^{ j -\frac{1}{2}} \simeq r_1^{ j -\frac{1}{2}}$ and $k_n (z)= K_{j-\frac{1}{2}+n-\alpha}(z)\simeq \frac{ \Gamma( \abs{j-\frac{1}{2}+n-\alpha })}{2}\qty( \frac{2}{z })^{\abs{j-\frac{1}{2}+n-\alpha }}  ~(n=0,1) $. In the limit $\abs{M} \ll M_
{PV} \ll 1/r_1 $, the ４-1 component $\sim \frac{1}{r_1^3}$ in the determinant is the most dominant, and the 3-2 component $\sim \frac{1}{r_1^2}$ is the secondary dominant. Then the determinant reduces to 
\begin{align}
    0= \frac{4\alpha^2}{ r_1^3 (j+1/2) } u_\frac{3}{2}  \frac{M_{PV}}{2} \frac{4\alpha}{r_1^2} v_{-\frac{1}{2}} \det \mqty( -2M_{PV} k_0( 2M_{PV} r_1)   & (M+E) k_0( \sqrt{M^2 -E^2}  r_1) \\  2M_{PV}  k_1(2M_{PV} r_1) & \sqrt{M^2 -E^2} k_1 (\sqrt{M^2 -E^2} r_1) ).
\end{align}
Thus the two solution on $r>r_1$ must be parallel at $r=r_1$ to obtain a continuum solution. Since $M_{PV} r_1 $ and $ \sqrt{M^2 -E^2}r_1 $ are small, $E$ is determined by
\begin{align}
    \frac{M+E}{\sqrt{M^2-E^2} } \qty(\frac{\sqrt{M^2-E^2}}{2M_{PV}})^{ \abs{j+\frac{1}{2}-\alpha }-\abs{j-\frac{1}{2}-\alpha } } +1=0 \label{eq:energy condition of localized mode}.
\end{align}
There exists a solution if and only if $M<0$ since $E^2< M^2$. We obtain the same condition for $j<0$, where the eigenfunction is approximated by
\begin{align}
    \psi_{i+}& \simeq \mqty( -\frac{2\alpha}{ M_{PV} r_1^2 }   {}_1 F_1(-j+\frac{1}{2}+\frac{1}{2}, -j+ \frac{3}{2}; \alpha \frac{r^2}{r_1^2}) e^{i(j-\frac{1}{2})\theta} \\ -2\frac{-j+\frac{1}{2}}{r} {}_1 F_1(-j+\frac{1}{2}+\frac{1}{2}, -j+ \frac{1}{2}; \alpha \frac{r^2}{r_1^2}) e^{i(j+\frac{1}{2})\theta} ) r^{ -j +\frac{1}{2}}e^{- \frac{ \alpha r^2}{ 2r_1^2}} \\
    D^2 \psi_{i+} & \simeq \mqty( \frac{2\alpha}{ M_{PV} r_1^2 } M_{PV}^2 {}_1 F_1(-j+\frac{1}{2}+\frac{1}{2}, -j+ \frac{3}{2}; \alpha \frac{r^2}{r_1^2}) e^{i(j-\frac{1}{2})\theta} \\ -2\frac{-j+\frac{1}{2}}{r} \frac{4\alpha}{r_1^2} {}_1 F_1(-j+\frac{1}{2}+\frac{1}{2}, -j+ \frac{1}{2}; \alpha \frac{r^2}{r_1^2}) e^{i(j+\frac{1}{2})\theta} ) r^{- j +\frac{1}{2}} e^{- \frac{ \alpha r^2}{ 2r_1^2}}
\end{align}
and the dependence of $r_1$ is the same as $ j>0$. 

$\abs{j+\frac{1}{2}-\alpha }-\abs{j-\frac{1}{2}-\alpha } $ takes three values
\begin{align}
    \abs{j+\frac{1}{2}-\alpha }-\abs{j-\frac{1}{2}-\alpha } =\left\{ \begin{array}{cc}
        +1 &  (j-\frac{1}{2}>\alpha) \\
        2(j-\alpha) & (j+\frac{1}{2} >\alpha> j-\frac{1}{2})  \\
        -1 & (j+\frac{1}{2}<\alpha)
    \end{array} \right. .
\end{align}
When $j -1/2> \alpha$ or $j+\frac{1}{2}<\alpha$, $E\sim M_{PV}$ so it is contradict with $E^2<M^2$. A solution of \eqref{eq:energy condition of localized mode} exists only when $j+\frac{1}{2} >\alpha> j-\frac{1}{2}$, i.e. $j= [\alpha]+\frac{1}{2}$. $E$ is an odd function of $\alpha-[\alpha]-1/2$ and described as the inverse function of
\begin{align}
    \alpha -[\alpha]= \frac{1}{2} \frac{ \log ( -\frac{M+E}{2 M_{PV} })}{ \log ( \frac{ \sqrt{M^2 -E^2 }}{2 M_{PV} }) }.
\end{align}
Thus $E$ is approximated as
\begin{align}
    E\simeq \left\{ \begin{array}{cc}
        -\abs{M} & ( \alpha -[\alpha] \sim 0 ) \\
        \abs{ 2M \log( \frac{\abs{M}}{2 M_{PV}}) }\qty( \alpha-[\alpha] -\frac{1}{2}) & ( \alpha -[\alpha] \sim \frac{1}{2} )   \\
        \abs{M} &  ( \alpha -[\alpha] \sim 1 )
    \end{array} \right. . 
\end{align}

\end{document}

%% file: main_introduction.tex
\section{Introduction}

Lattice gauge theory has played an essential role in theoretical computations of hadronic processes. It is usually formulated on a square lattice with periodic or anti-periodic boundary conditions, to regularize the quantum field theory to a well-defined integral of finite degrees of freedom. The systematics due to finite lattice spacing and finite volume size is well understood and the continuum limit can be taken in a controlled manner.

With a gravitational background, on the other hand, lattice field theory is less developed, compared to the case in a flat space time. In order to describe the nontrivial curved metric, one needs to modify the structure of the lattice itself, which has been often studied by triangular lattices \cite{Hamber2009Quantum,Regge1961general,brower2016quantum,AMBJORN2001347Dynamicallytriangulating,Brower2017LatticeDirac,Catterall2018Topological,ambjorn2022topology,Brower:2019kyh,Brunekreef:2021iay,Asaduzzaman:2022kxz,Ambjorn:2022naa}. The continuum limit is, however, nontrivial as it is difficult to control the number of sites, links, and their angles and lengths at the same time. The recovery of the symmetry that the target continuum theory has is not obvious, either \cite{Brower2015Quantum}. 

In our previous work \cite{Aoki:2022cwgCurved}, we proposed a novel formulation of curved space field theory using a fermion system on a square lattice. As a mathematical base, we rely on Nash's embedding theorem \cite{Nash1956TheImbedding,Gromov1970Embeddingsand} which states that any curved Riemannian manifold can be isometrically embedded into a finite-dimensional flat Euclidean space and their metric and vielbein are uniquely determined up to gauge transformations. In our formulation a square lattice is used to regularize the flat higher dimensional space, and the embedding is realized by a curved domain-wall in the fermion mass term. In the same way as in the flat domain-wall fermion \cite{KAPLAN1992342AMethod,Shamir1993Chiral,Furman1995Axial} the massless edge-localized modes appear on the domain-wall, and they feel ``gravity" through the Einstein's equivalence principle. Note that we do not assign any link or site variables to describe gravitational degrees of freedom. They appear as effective low-energy fields. 

In the analysis of the circle $S^1$ domain-wall fermion on a two-dimensional square lattice as well as the $S^2$-shaped curved domain-wall fermion in three dimensions, we found that 1) chiral edge-localized modes appear at the domain-wall, 2) they feel gravity through the induced spin connection, and 3) the continuum extrapolation including the rotational symmetry is well controlled, tuning the lattice spacing of the higher dimensional square lattice.

A similar approach to our study was already discussed in condensed matter physics. In \cite{Imura2012Spherical,Takane2013UnifiedDescription}, they found in the continuum effective theory of a topological insulator that the edge-localized modes on a spherical curved surface are described by the induced spin connections. Our lattice result below agrees well with their analytic observation. In \cite{Hotta:2013qta,Hotta:2022aiv},  it was proposed how to demonstrate the inflation mechanism using the expanding edge of the quantum Hall systems. In fact, our study has triggered a collaboration \cite{AFKKMinpre} with condensed matter physicists for microscopically understanding the Witten effect \cite{Witten:1979Dyons}, which is a theoretical prediction that a magnetic monopole becomes a dyon inside topological insulators.

In this work, in constant to our previous work \cite{Aoki:2022cwgCurved} where we studied a free fermion system only, we introduce nontrivial $U(1)$ link variables to investigate the anomaly inflow between the bulk and edge. In continuum theory, it is known that the time-reversal symmetry anomaly \cite{Alvarez-Gaume:1984zst} of the bulk and edge modes is canceled through the Atiyah-Patodi-Singer index theorem \cite{Atiyah1975spectral,Witten:2015Fermion}, to which a mathematical relation with the domain-wall fermion Dirac operator was proved \cite{Fukaya_2017Atiyah-Patodi-Singer, fukayaFuruta2020physicistfriendly}.

With a weak and smooth $U(1)$ flux in the $S^1$ domain-wall fermion, we will show that the $\eta$ invariant of the massive Dirac operator is consistent with the APS index on a disk. As far as we know, this is the first example of a nontrivial APS index realized on a lattice except for the flat two or four-dimensional tori \cite{FukayaKawai2020TheAPS}.

We will also study the cases where the gauge field has a singular defect: a vortex in two dimensions \cite{Lee:2019rfbFractionalchargebound,Khalilov:2014rka} and a magnetic monopole \cite{Dirac:1931kp} in three dimensions \cite{Yamagishi:1982wpTHE,Yamagishi1983Fermion-monopole,Yamagishi1984Magnetic,WU1976365}. For these strong field cases, the Wilson term in the fermion action plays a key role: it dynamically creates another domain-wall near the defects. The creation/annihilation of the domain-walls corresponds to topology changes of the total system, but the anomaly inflow is still consistent with the cobordism. If we consider the Dirac operator as a Hamiltonian in one dimension higher system \cite{Imura2012Spherical,Takane2013UnifiedDescription,Rosenberg:2010iaWitteneffect,Qi:2008ewTopologicalFieldTheory}, the creation of the domain-wall naturally explains Witten effect \cite{Witten:1979Dyons, Yamamoto2020Magneticmonopoles,Rosenberg:2010iaWitteneffect,Qi:2008ewTopologicalFieldTheory} which indicates that the defects are electrically charged inside topological insulators.

The rest of the paper is organized as follows. In Sec. \ref{sec:review}, we explain our lattice setup for curved domain-wall systems. In Sec. \ref{sec:S1_AnomalyInflow}, we embed $S^1$ domain-wall into a square lattice and assign a weak and uniform $U(1)$ connection inside the wall. We analyze how the connection affects the dynamics of edge modes and how the anomaly inflow is detected. In Sec. \ref{sec:S1_Witten}, we squeeze the $U(1)$ flux in a single plaquette. We show an extra domain-wall is created by a Wilson term and localized mode appears at the plaquette. In Sec. \ref{sec:S2_Witten}, we confirm a similar bound state around a magnetic monopole in the $S^2$ domain-wall fermion system.

~

%% file: main_review.tex
\section{Curved domain-wall fermion on a lattice}
\label{sec:review}

In our previous work \cite{Aoki:2022cwgCurved}, we investigated free fermion systems on  two and three-dimensional square lattices having a curved domain-wall in their mass term. We found that the edge-localized modes appear at the spherical ($S^1$ and $S^2$) domain-walls \cite{Imura2012Spherical,Takane2013UnifiedDescription}, the effect of gravity or induced Spin and Spin$^c$ connections encoded in the Dirac spectrum, and a monotonic scaling behavior towards the continuum limits including recovery of the rotational symmetry.

The key mathematical ideas in our previous work are the Nash's embedding theorem \cite{Nash1956TheImbedding} and Einstein's equivalence principle. The Nash's theorem assures that any curved Riemannian manifold can be isometrically embedded into a finite-dimensional flat Euclidean space and its metric and vielbein are uniquely induced by the embedding function. When the motion of some particles or fields are constrained to the embedded manifold, they feel gravity, by the equivalence principle, through the induced metric or Spin and Spin$^c$ connections.

In our previous work \cite{Aoki:2022cwgCurved}, we regularized the higher-dimensional Euclidean space on a square lattice, and realized the embedding function by the domain-wall mass term in the fermion. It this work, we investigate the effect of electro-magnetic gauge fields focusing on the anomaly. On a higher-dimensional square lattices, it is straightforward to introduce the gauge fields by the standard link variables and associated covariant difference operators.

We denote a $(n+1)$-dimensional periodic square lattice by $(\mathbb{Z}/N\mathbb{Z})^{n+1}$. The physical lattice size is $L=N/a$ with the lattice spacing $a$. The lattice site coordinate is given by $x_i=a\hat{x}_i$, where $ \hat{x}_i$ takes a　half-integer value in the range $-(N-1)/2 , -(N-1)/2+1, \cdots,   (N-1)/2 $\footnote{Here we set the origin at the center of $n+1$-dimensional hypercube.}. The covariant difference operators in the $i$-th direction $\nabla_i$ and its adjoint $\nabla_i^\dagger$ are given by
\begin{align}
(\nabla_\mu \psi)_{x} &=U_\mu (x)\psi_{x+ a \hat{\mu}}-\psi_{x}\\
(\nabla_\mu ^\dagger \psi)_{x} &=U_\mu^\dagger (x-a \hat{\mu}) \psi_{x- a \hat{\mu}}-\psi_{x},
\end{align} 
where $U_\mu (x)$ is a link variable and defined by
\begin{align}
    U_\mu (x)=P\exp(i\int_{x+a \hat{\mu}  }^x  A  ).
\end{align}
$A$ is a gauge field. Then the hermitian domain-wall fermion Dirac operator is given by
\begin{align} \label{eq:Dirac op on lattice general}
H&=\frac{\bar{\gamma}}{a}\qty( \sum_{i=1}^{n+1} \qty[\gamma^i
\frac{\nabla_i-\nabla^\dagger_i}{2} +\frac{1}{2}\nabla_i
\nabla^\dagger_i]+  \text{sign}(f)ma),
\end{align}
where $f: \mathbb{R}^{n+1} \to \mathbb{R}$ is a real-valued smooth function and generates a domain-wall $Y=\set{x \in \mathbb{R}^{n+1} |f(x)=0 }$. Here we take $m$ to be positive and the Wilson term to be unity. We denote the Dirac's gamma matrices by
\begin{align}
    \gamma^a=-\sigma_2 \otimes \tilde{\gamma}^a,\
\gamma^{n+1}=\sigma_1 \otimes 1 ,\ \bar{\gamma}=\sigma_3 \otimes 1,
\label{eq:expression of gamma matrix general}
\end{align}
where $\tilde{\gamma}^a~(a=1,\cdots n)$ are the $2^{[n/2]}\times 2^{[n/2]}$ gamma matrices ($[\alpha]$ denotes the Gauss symbol or the integer part of $\alpha$) which satisfy the Clifford algebra in $n$ dimensions.

Note that the Pauli matrices $\sigma_{1,2,3}$ operate as flavor matrices rather than spinors when $n$ is even. The doubling of the flavor space is introduced to make the massive Dirac operator Hermitian with the $ \mathbb{Z}_2$ grading operator $\bar{\gamma}$.

In the continuum limit $a\to 0$, this operator converges to
\begin{align}
    H=\bar{\gamma} \qty( \sum_{I=1}^{n+1} \gamma^I \qty(\pdv{}{x^I} -iA_I)+m
\text{sign}(f)).
\end{align}
In our previous work, we compared our numerical result for the lattice Dirac eigenvalues and those analytically obtained in continuum to find the footprint of gravity in the spectrum of the edge states localized at $S^1$ or $S^2$ domain-walls.
In this work, we extend the analysis to the system with nontrivial $U(1)$ link gauge fields and investigate the anomaly inflow between bulk and edge fermion modes.

%% file: main_S1_weakU1b.tex
\section{Anomaly of the $S^1$ domain-wall fermion}
\label{sec:S1_AnomalyInflow}

In this section, we consider an $S^1$ domain-wall fermion coupled to $U(1)$ gauge field on a two-dimensional square lattice. The gauge field changes the eigenvalues of the edge-localized modes, which yields the anomaly of the time reversal ($T$) symmetry \cite{Alvarez-Gaume:1984zst}. As the whole two-dimensional fermion system respects the $T$ symmetry, this $T$ anomaly of the edge modes must be canceled by the bulk modes. We show how this $T$ anomaly correspondence between the bulk and edge is protected in the curved domain-wall system, and its consistency with the Atiyah-Patodi-Singer index on the disk whose boundary is located at the $S^1$ domain-wall \cite{Witten:2015Fermion}. 

First let us consider a weak and uniform $U(1)$ $1$-form gauge field in continuum theory in flat two dimensions, given by the vector potential
\begin{align} \label{eq:U1 flux on S1}
    A= \left\{ \begin{array}{ll}
        \alpha \qty(-\frac{y}{r_1^2} dx+ \frac{x}{r_1^2} dy)=\alpha
\frac{r^2}{r_1^2}d\theta & (r<r_1) \\
    \alpha \qty(-\frac{y}{r^2} dx+ \frac{x}{r^2} dy) =\alpha d\theta& (r>r_1)
    \end{array} \right. ,
\end{align}
whose field strength is $F_{12}=2\alpha/r_1^2$ for $r<r_1$ and zero for $r>r_1$. $\alpha$ represents the total flux (divided by $2\pi$) of the entire system.

It is straightforward to translate this continuum vector potential into the link variables on the lattice. The corresponding covariant difference operator in the $x$-direction is given by (denoting $x=a\hat{x}$ and $y=a\hat{y}$)
\begin{align}
    \nabla_1 \psi_{(\hat{x},\hat{y})} =\exp(i \int_
{(\hat{x}+1,\hat{y}) }^{(\hat{x},\hat{y}) }A )
\psi_{(\hat{x}+1,\hat{y}) } -\psi_{(\hat{x},\hat{y}) },
\end{align}
and that in $y$-direction is defined by the same manner. It is obvious that this operator converges to $ \partial_i -i A_i$ in the continuum limit. In order to achieve the periodic boundary condition on the fermion, we set the link variables connecting $x=N/2$ and $-N/2$  for all $y$ and those connecting $y=N/2$ and $-N/2$ for all $x$ to unity. The field strength around the boundary gives little effect on the edge-mode spectrum.

\begin{figure}[h]
    \centering
   \includegraphics[width=\textwidth]{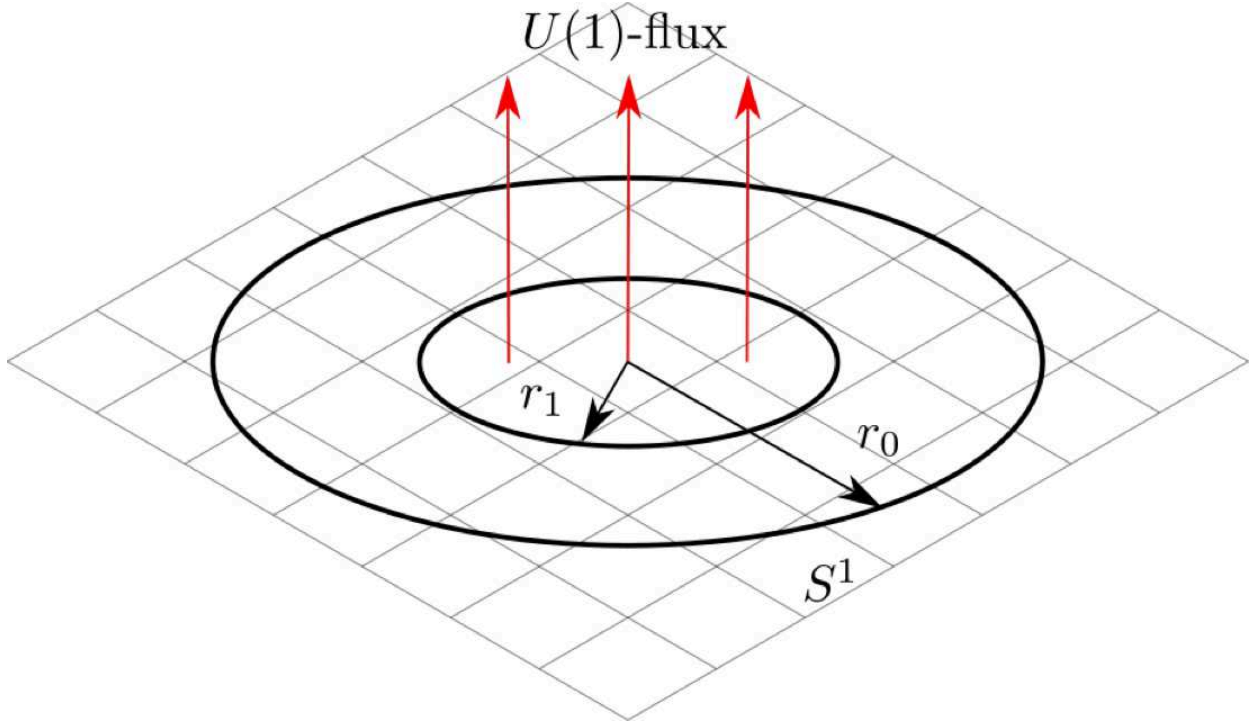}
    \caption{Our two-dimensional square lattice set up. The outer circle represents the $S^1$ domain-wall with radius $r_0$ and inside the inner circle with radius $r_1$ we put the uniform $U(1)$ gauge field strength.}
    \label{fig:U1 flux}
\end{figure}

With these link variables, we numerically solve the eigenproblem of the hermitian Dirac operator
\begin{align}\label{eq:Hermitian Wilson Dirac op of S1 in R2}
        H =\frac{1}{a}\sigma_3
\qty(\sum_{i=1,2}\qty[\sigma_i\frac{\nabla_i-\nabla^\dagger_i}{2}
+\frac{1}{2}\nabla_i \nabla^\dagger_i ]+a\epsilon m  ), 
\end{align}
where $\sigma_i$ are the Pauli matrices, and we assign the $S^1$ domain-wall with the radius $r_0>r_1$ by the step function
\begin{align}
    \epsilon( \hat{x}, \hat{y})=\left\{ \begin{array}{cc}
        -1 & (a \hat{r} <r_0 )  \\
        1 & ( a \hat{r} >r_0)
        \end{array}
        \right. .
\end{align}
We illustrate the lattice system in Fig.~\ref{fig:U1 flux}.

If we regard this two-dimensional system as a layer or sheet in three space dimensions, the $U(1)$ flux corresponds to a magnetic field penetrating the sheet in the range $r<r_1$. For the fermion modes localized at $r=r_0$, the magnetic field does not directly interact, but the Aharonov-Bohm effect \cite{Aharonov:1959fk} changes the spectrum of $H$.

In order to test the essential property of the edge-localized modes, we also numerically evaluate the expectation value of the gamma matrix in the normal direction of the $S^1$ domain-wall
\begin{align}
    \gamma_{\text{normal}}=\frac{x}{r} \sigma_1+\frac{y}{r} \sigma_2,
\end{align}
which is well-defined when $N$ is even. We call $ \gamma_{\text{normal}}$ the ``chirality" operator by analogy from the standard flat domain-wall fermion in five dimensions, although in our two-dimensional case there is already a more legitimate operator $\bar{\gamma}=\sigma_3$.

\begin{figure}[h]
    \centering
    \includegraphics[scale=1]{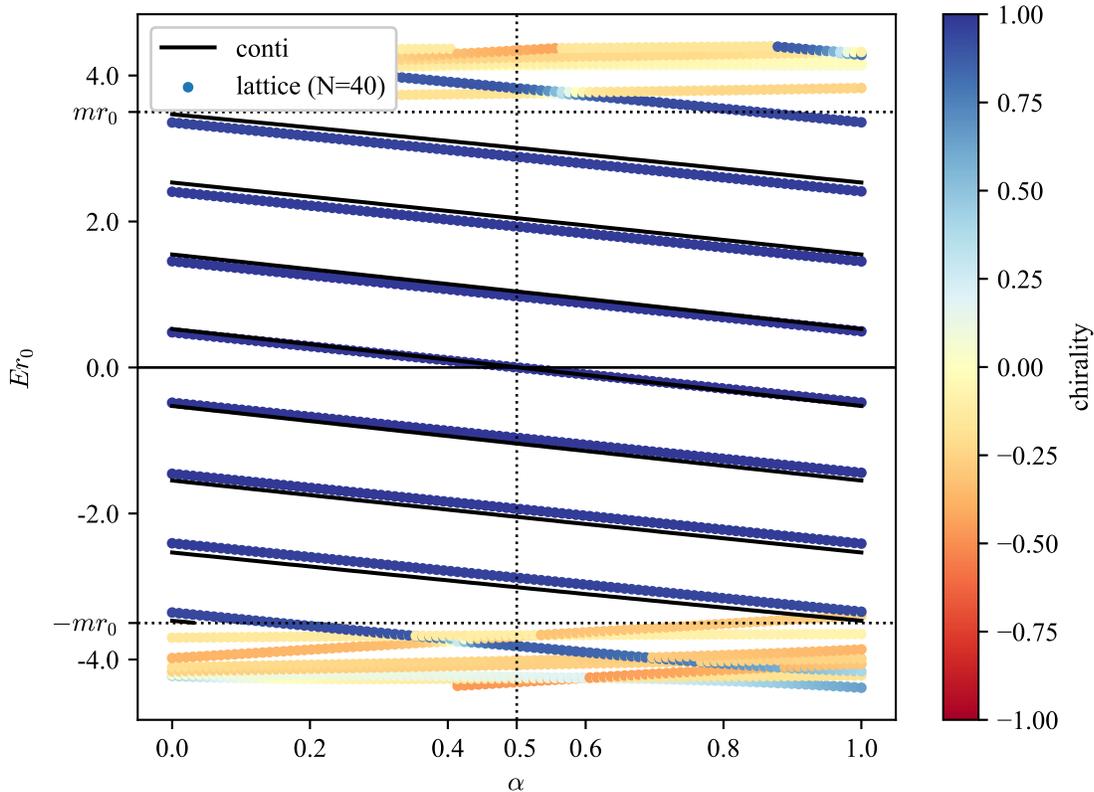}
    \caption{The eigenvalue spectrum of $H$ at $m= 14/L$, $r_0/a = 10$, $r_1/a=5$ and $L/a = 40$. The filled circles show the lattice data and the solid lines are the continuum prediction in Eq.~(\ref{eval-cont}). The color gradation of the data points from deep blue through light yellow to deep red represents the expectation value of the ``chirality" operator $\gamma_{\text{normal}}$.
}
    \label{fig:Anomaly inflow r1 =r0/2}
\end{figure}

In Fig.~\ref{fig:Anomaly inflow r1 =r0/2}, we plot the lattice results for the eigenvalues of $H$ as a function of $\alpha$. Here we set $m=14/L$, $L=40a$, $r_0=\frac{L}{4}$ and $r_1=\frac{r_0}{2}$. We can clearly see that the spectrum is monotonically decreasing linearly with $\alpha$.

The linear dependence of the spectrum is due to the Aharonov-Bohm effect. In the large $m$ limit in the continuum theory, we can analytically solve the eigenproblem of $H$ (see Appendix~\ref{sec:Eigenvalue of Edge mode S1}). The effective Dirac operator on the edge-localized modes reads
\begin{align}
    i r_0\Slash{D}^{S^1}=-i\pdv{}{\theta}+\frac{1}{2}-\alpha,
\end{align}
where $\theta$ denotes the coordinate of the $S^1$ in units of the radius $r_0$. $1/2$ in the second term is the induced $Spin^c$ connection and $\alpha$ originates from the Aharonov-Bohm phase of the wave function. The eigenvalues are then obtained as
\begin{align}
\label{eval-cont}
E_n r_0 = n+\frac{1}{2}-\alpha &~ (n\in \mathbb{Z}).
\end{align}
Note that the effect of the $Spin^c$ connection $1/2$ is canceled by the Aharonov-Bohm effect at $\alpha=1/2$. At $\alpha=1$ the spectrum comes back to the original distribution at $\alpha=0$ but each eigenvalue shifts from the level $n$ to the next lower level $n-1$. This behavior is similar to the Thouless charge pump \cite{Thouless1983Quantization}. This continuum prediction presented in Fig.~\ref{fig:Anomaly inflow r1 =r0/2} by the solid lines agrees well with the lattice data.

\begin{figure}[h]
    \centering
    \includegraphics[width=\textwidth]{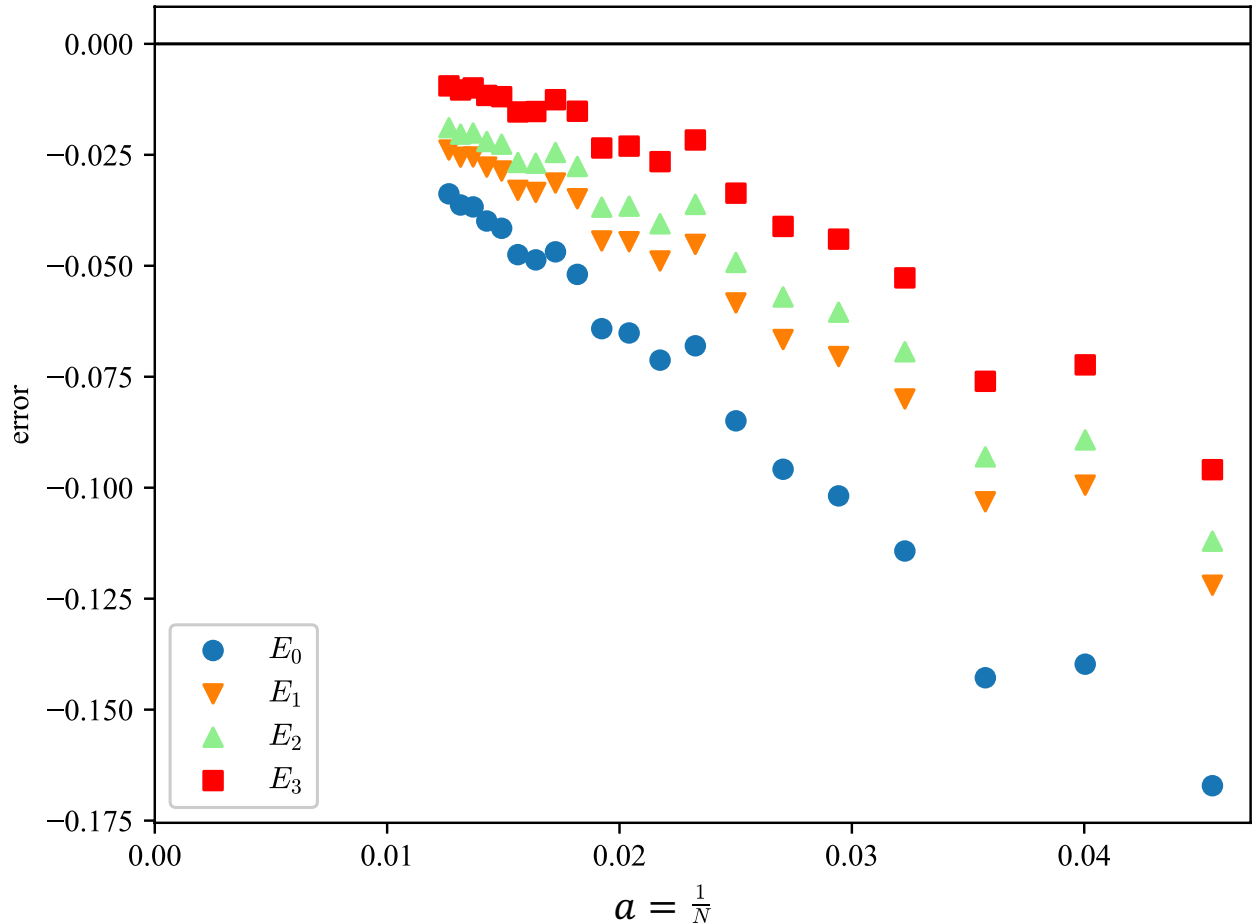}
  \caption{The relative deviation of the eigenvalue $\qty(E_{n}-E_n^\text{cont.})/E_n^\text{cont.}$, where $E_n^\text{cont.}$ denotes the continuum prediction. 
  }
  \label{fig:continuum limit of error}
\end{figure}

The cut-off dependence of the relative eigenvalue difference for $E_i(i=0,1,2,3)$ at $\alpha=0.25$ is plotted in Fig.~\ref{fig:continuum limit of error}, which shows a monotonous decrease towards the continuum limit.

\begin{figure}
    \centering
    \includegraphics[width=\textwidth]{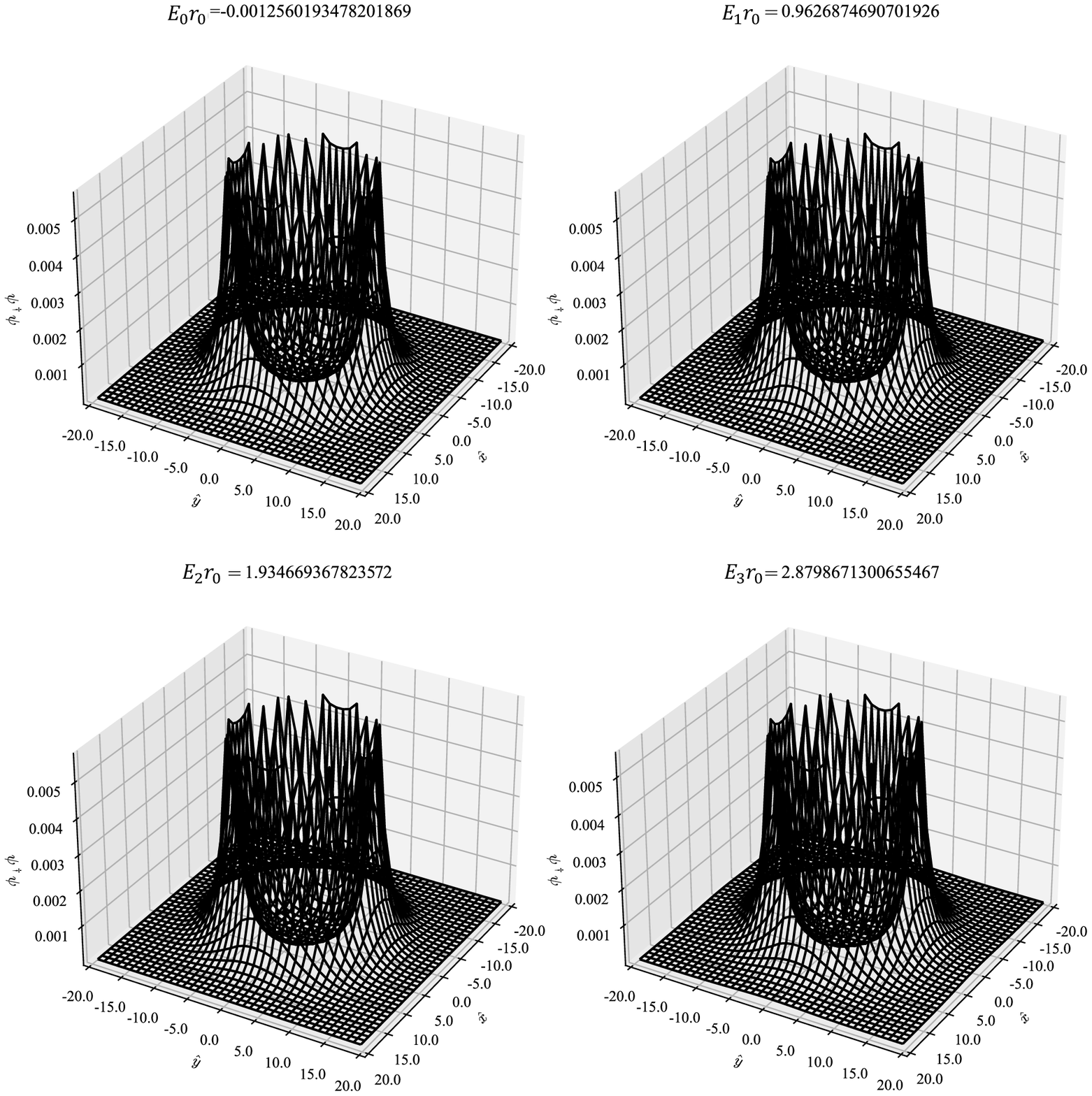}
    \caption{The amplitude of the edge modes with $E_{0\mathchar`-3} $ at $\alpha=0.5$, $m= 14/L$, $r_0/a = 10$, $r_1/a=5$ and $L/a = 40$.} 
    \label{fig:Edgemode S1}
\end{figure}

As the color gradation of the data points from the dark red ($\gamma=-1$) to dark blue ($\gamma=+1$) shows, we find that the eigenmodes with $\abs{E}<m$ are almost ``chiral" or $\gamma_{\text{normal}}\sim +1$.
We also find that these modes are well localized at $r=r_0$ as Fig.~\ref{fig:Edgemode S1} indicates. 

The linear shift of the Dirac eigenvalues due to $\alpha$ reflects the fact that the edge massless Dirac fermion suffers from the anomaly of the $T$ symmetry. The asymmetry of the spectrum is measured by the Atiyah-Patodi-Singer $\eta$ invariant \cite{Atiyah1975spectral}
\begin{align}
    \eta( i\Slash{D}^{S^1})=&\lim_{\epsilon \to 0} \sum_{\lambda \neq
0} \frac{\lambda}{ \abs{\lambda}^{1+\epsilon}}+ \dim \text{Ker}(
i\Slash{D}^{S^1})
\sum_{\lambda=0} 1,
\end{align}
where $\lambda$ are the eigenvalues of $i\Slash{D}^{S^1}$. We can directly confirm with the Pauli-Villars regularization that the phase of the fermion determinant is the $\eta$ invariant:
\begin{align}
   \lim_{\mu\to \infty} \det
\frac{\Slash{D}^{S^1}}{\Slash{D}^{S^1}+\mu}\sim \prod_{\lambda}
\frac{-i\lambda}{\mu}\propto \exp(-i\frac{\pi}{2}
\eta(i\Slash{D}^{S^1})). 
\end{align}
Since the sign of each eigenvalue flips under $T$ transformation, the edge mode partition function breaks the $T$ symmetry when $\eta(i\Slash{D}^{S^1})/2$ is a non-integer, even when the fermion action is $T$ invariant.

However, for the total two-dimensional lattice Dirac operator, $\eta(H)$ is always an even integer, and $\det H$ is real. This indicates that the above $T$ anomaly of the edge modes is canceled by the bulk contribution so that the $T$ symmetry of the total system is protected. In Refs.~\cite{Fukaya_2017Atiyah-Patodi-Singer,fukayaFuruta2020physicistfriendly} it was mathematically proved that for any domain-wall fermion Dirac operator $H$ on a closed Riemannian manifold $X$, $-\eta(H)/2$ is guaranteed to be an integer since it is equal to the Atiyah-Patodi-Singer index on a submanifold $X_-$, or the negative region of the fermion mass. In \cite{FukayaKawai2020TheAPS}, it was shown on a square lattice with a flat domain-wall, $-\eta(H)/2$ is consistent with the APS index, assuming that the link variables are sufficiently smooth.

In our two-dimensional curved domain-wall fermion system, we can analytically estimate the edge mode contribution as
\begin{align}
    -\frac{1}{2}\eta( i\Slash{D}^{S^1} )=& -\frac{1}{2}\qty(
\lim_{\epsilon\to 0} \sum_{n\in \mathbb{Z}}
\frac{n+\frac{1}{2}-\alpha}{\abs{n+\frac{1}{2}-\alpha}^{1+\epsilon} }
+\#\set{ \text{zero modes}} ) \nonumber\\
    =&[\alpha +\frac{1}{2}] -\alpha,
\end{align}
where $[A]$ denotes the Gauss symbol taking the integer value of $A$. Since it is obvious that the total $\eta$ invariant is $-\eta(H)/2 = [\alpha +\frac{1}{2}] $, the bulk contribution must be $+\alpha$, which agrees well with the perturbative evaluation \cite{FukayaKawai2020TheAPS} 
\begin{align}
   -\frac{1}{2}\eta(H)^{\text bulk} =  \frac{1}{2\pi} \int_{r<r_1} F=
\frac{1}{2\pi} \int_{S^1} A=\alpha.
\end{align}
This observation indicates that the $T$ anomaly matching between bulk and edge is well described through the $\eta$ invariant of $H$ or the APS index on our square lattice even when the domain-wall is curved.

As far as we know, this is the first example of a nontrivial APS index on a curved manifold realized in lattice gauge theory setup.

%% file: main_WittenEffect.tex
\section{Topology change of the domain-wall and Witten effect}
\label{sec:S1_Witten}

In the previous section, a uniform weak gauge field background has been considered. We have confirmed the $T$ anomaly inflow on a square lattice, where the asymmetry of the edge localized mode's spectrum is
compensated by the bulk contribution, and the $T$ symmetry is protected as a consequence of the APS index theorem on a two-dimensional disk. In this section, we consider a strong gauge field concentrated in a short range, which yields a more dynamical change to the fermion system.

\subsection{Topology changes: creation and annhilation of the domain-wall}

Let us take a very small value of $r_1$, while the total $U(1)$ flux is unchanged. Then, at $r_1<a/2$, only one plaquette at the origin gives $-1$, representing a strong gauge field. With such a high energy configuration in a short range, it is nontrivial if the anomaly inflow persists via the APS index theorem. As is shown below, the anomaly matching is still valid, but in a drastically different way from the weak field case in the previous section.

\begin{figure}
    \centering
    \includegraphics[scale=0.8]{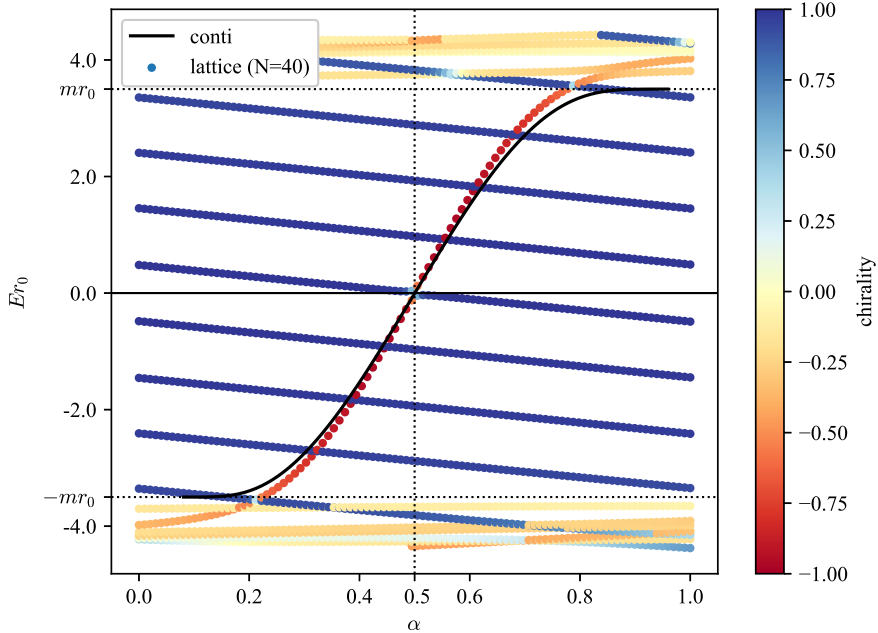}
    \caption{The eigenvalue spectrum of \eqref{eq:Hermitian Wilson Dirac op of S1 in R2} $m = 14/L$, $\hat{r}_0=r_0/a = 10$, $\hat{r}_1=0.001$ and $L = 40a$. The filled circles show the lattice data and the solid lines are the continuum prediction.}
    \label{fig:Anomaly inflow r1 =0.001*a}
\end{figure}

Figure~\ref{fig:Anomaly inflow r1 =0.001*a} shows the $\alpha$ dependence of the Dirac eigenvalue spectrum. The uniform linear shift with $\alpha$ of the edge-localized modes with positive chirality is seen as in the weak field case. However, a remarkable difference is that one negative chirality mode appears when $\alpha=0.2$  at energy level $\sim -m$, crosses zero at $\alpha=1/2$ and disappears at the level $\sim m$.

\begin{figure}
	\begin{minipage}[b]{\textwidth}
		\centering
		\includegraphics[scale=0.8]{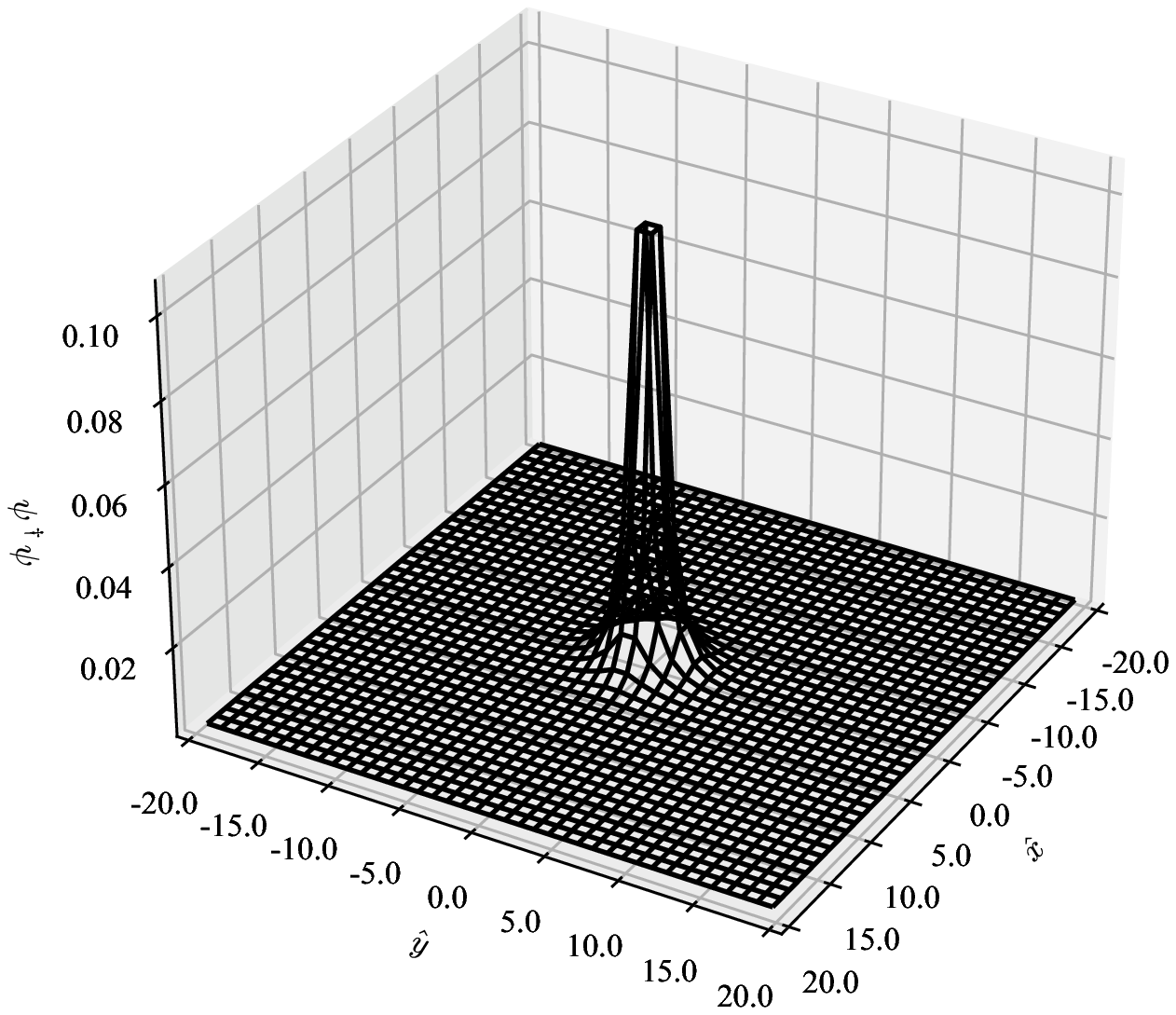}
		\caption{The amplitude of the eigenstate localized at the origin when $\alpha=0.4$, $m = 14/L$, $r_0/a = 10$, $r_1<a/2$ and $L/a = 40$.}
		\label{fig:localized mode A=0.4}
	\end{minipage}
	
	
	\begin{minipage}[b]{\textwidth}
		\centering
		\includegraphics[bb=118 226 478 610,scale=0.8]{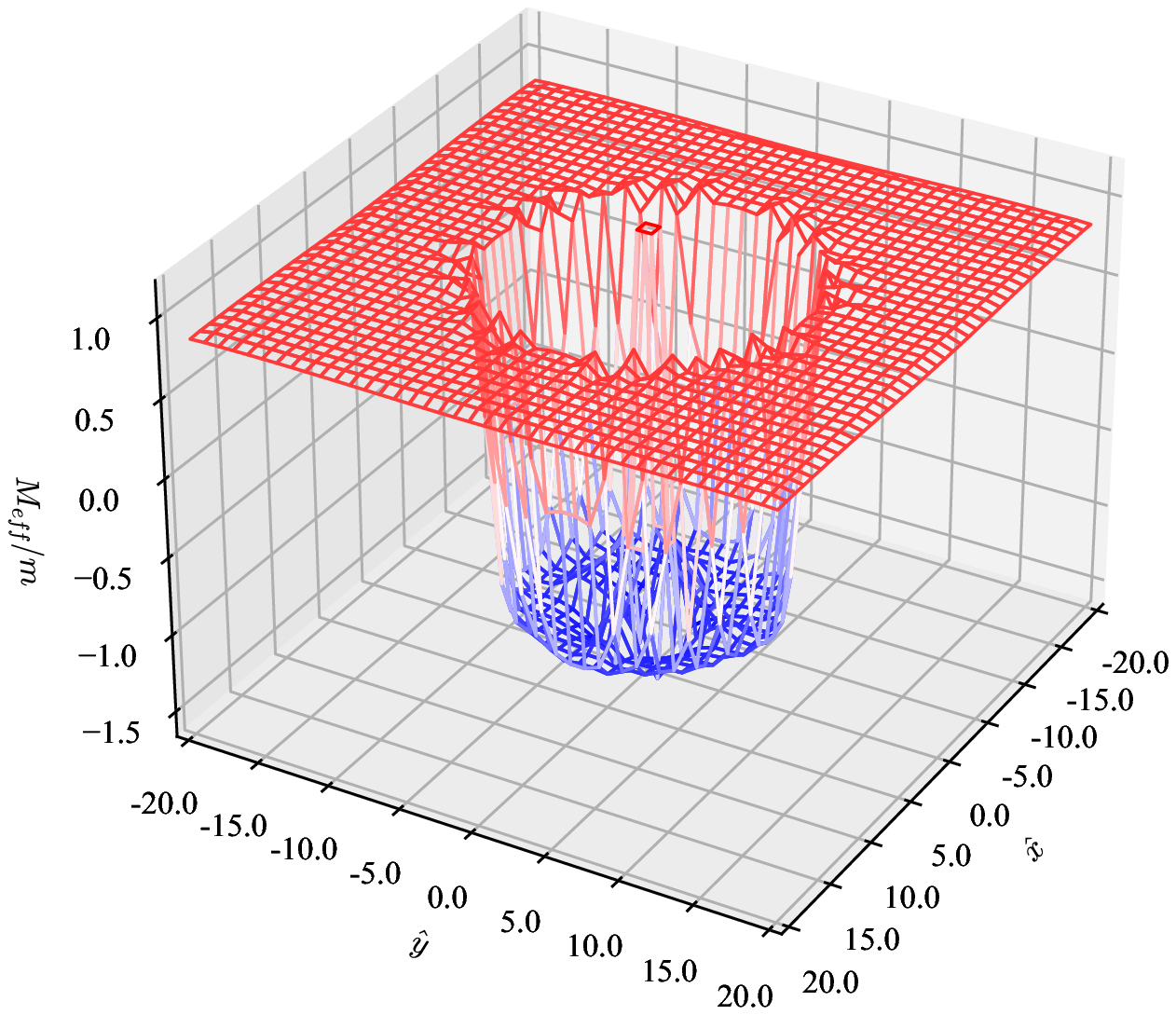}
		\caption{The ratio of the effective mass $M_{eff}$ and $m=\abs{M}$ when $m = 14/L$, $\hat{r}_0=r_0/a = 10$, $\hat{r}_1=0.001$ and $L = 40 a$.}
		\label{fig:effective mass term}
	\end{minipage}
\end{figure}

In fact, as presented in Fig.~\ref{fig:localized mode A=0.4} this mode is localized at the origin, where we do not assign any domain-wall structure in the mass term but the strong $U(1)$ gauge field is located. Since there was no such mode in the weak field case in the previous section, it is natural to assume that this mode is somehow excited by the strong gauge field potential.

We numerically investigate this issue and find that the Wilson term plays a key rule. As is well known, the Wilson term gives additive mass renormalization due to the violation of the chiral symmetry. In lattice QCD it is known that the stronger the coupling is, the larger the additive mass shift becomes in the positive direction. We confirm that the same mass shift happens in our system but only locally near the origin where the strong background is given. In Fig.~\ref{fig:effective mass term}, we plot the local expectation value of the ``effective" mass 
\begin{align}
   M_{eff}= \frac{1}{ \psi(x)^\dagger \psi(x)}  \psi(x)^\dagger
\qty(\epsilon m+\sum_{i=1,2}\frac{1}{2a}\nabla_i \nabla^\dagger_i  ) \psi(x),
\end{align}
where the second term contains the dependence on the link variables. The renormalized mass near the origin becomes positive, creating another domain-wall around it. We can identify this as dynamical creation of the domain-wall and the center-localized mode appears as the new edge mode on it.

In terms of the APS index expressed by the $\eta$ invariant of $H$, the topology of the target spacetime defined by the negative mass region, is changed from a disk with one $S^1$ boundary to a cylinder with two $S^1$ boudaries at $r=r_0$ and $r=r_1$. Thus, the $U(1)$ gauge field can make creation/annihilation of the domain-walls and topology changes of the bulk manifold on which the APS index describes the anomaly inflow.

In particular, it is interesting to see the case $\alpha=0.5$, where two zero modes appear: one is on the original edge at $r=r_0$ and another center-localized mode (to be precise, they are split due to the interference, which will be discussed later in detail). In this case, the Dirac operator becomes real and the number of zero modes is known to be the mod-two index \cite{AtiyahSinger1971TheIndex5}, exhibiting a global anomaly \cite{Witten:1982fp}. On $S^1$, the mod-two index, being a cobordism invariant, must be zero when it is boundary of a disk. Existence of the two zero modes can be understood as a new type of anomaly inflow: changing topology from disk to cylinder by the dynamical creation of the domain-wall, another zero-mode at the center appears in order to make the total mod-two index zero (mod 2) and the system free from global anomaly.

In the above discussion on the dynamical creation of the domain-wall, the Wilson term, which represents a higher derivative term in the continuum limit, is essential. It is interesting to ask if we can understand the center-localized mode in continuum theory, too. We find a positive answer to this question but the Dirac equation requires a second derivative term with $M=-m$ which can be understood as a Pauli-Villars contribution.

The details of the computation are given in Appendix~\ref{sec:Eigenvalue of localized mode S1}. We confirm that for $\abs{E}<m$ only one solution exists and the obtained $E$ as a function of $\alpha$ is plotted in Fig.~\ref{fig:Anomaly inflow r1 =0.001*a}, where agrees well with our lattice data.

\subsection{Microscopic description of Witten effect}

The appearance of the center-localized fermion can be viewed as an electric ``charge" of the electro-magnetic defect. It was shown by Witten \cite{Witten:1979Dyons} that a monopole in a topological insulator, or equivalently in the $\theta=0$ vacuum, obtains a half electric charge. Although our object is not a monopole but a vortex, let us call the phenomenon that some defect in the electro-magnetic field obtains a fractional charge \cite{Lee:2019rfbFractionalchargebound,Khalilov:2014rkaBoundstates,Rosenberg:2010iaWitteneffect,Qi:2008ewTopologicalFieldTheory,Zhao2012Amagneticmonopole} the ``Witten effect" since we believe that their microscopic mechanism is essentially the same, as explained below\footnote{We will discuss the monopole case in the next section.}.

To this end, let us switch the view from two-dimensional Euclidean spacetime to two-dimensional space with the Hamiltonian given by (\ref{eq:Hermitian Wilson Dirac op of S1 in R2}). The negative mass region with $r<r_0$ is regarded to be a topological insulator, while outside is normal insulator. We can regard the $U(1)$ flux as a strong magnetic field penetrating the sheet.

\begin{figure}
    \centering
    \includegraphics{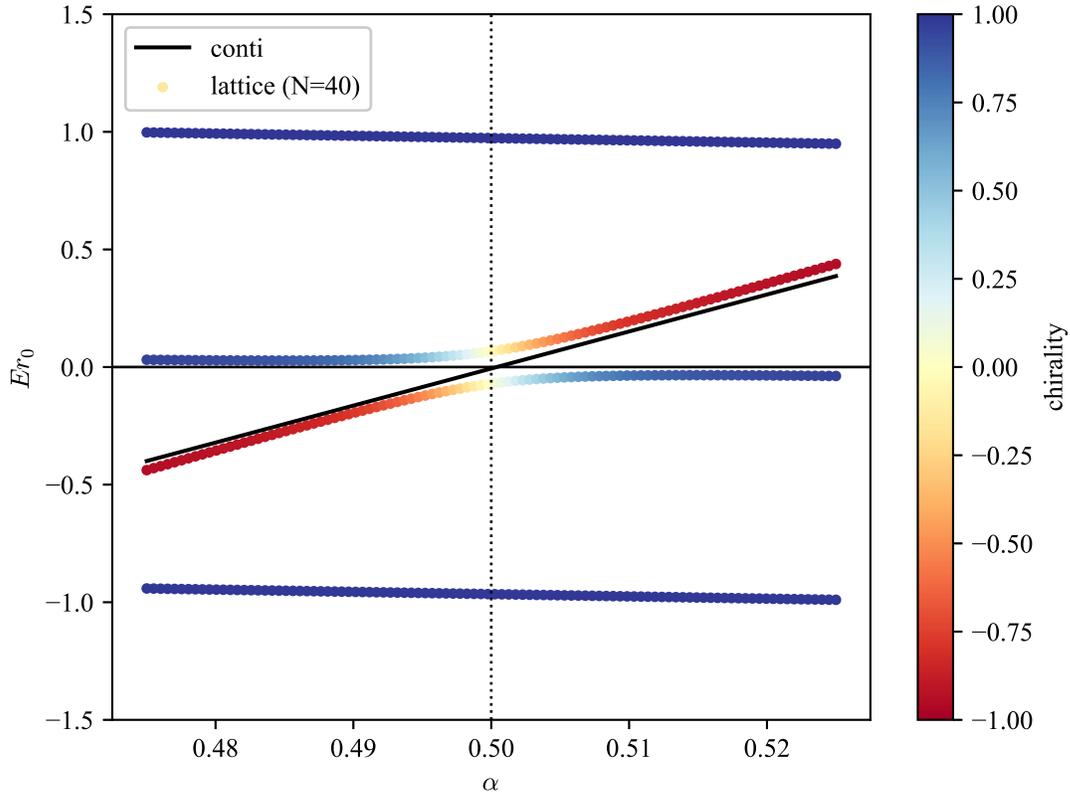}
    \caption{Enlarged view of the $\alpha=0.5$ of Fig.~\ref{fig:Anomaly inflow r1 =r0/2}}
    \label{fig:anomaly inflow enlarge}
\end{figure}

\begin{figure}
  \begin{minipage}[b]{0.5\linewidth}
    \centering
    \includegraphics[width=\textwidth]{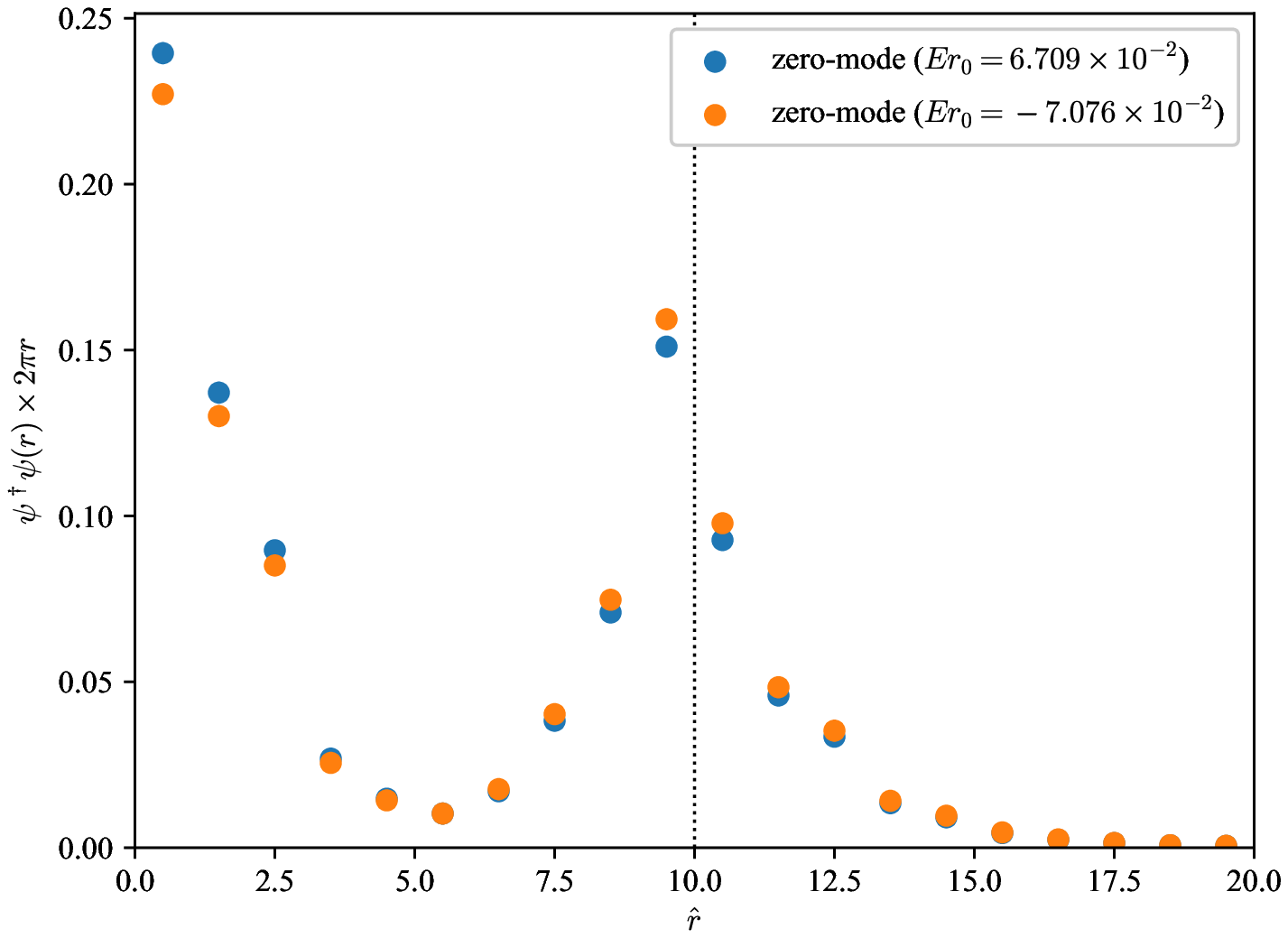}
  \end{minipage}
  \begin{minipage}[b]{0.5\linewidth}
    \centering
    \includegraphics[width=\textwidth]{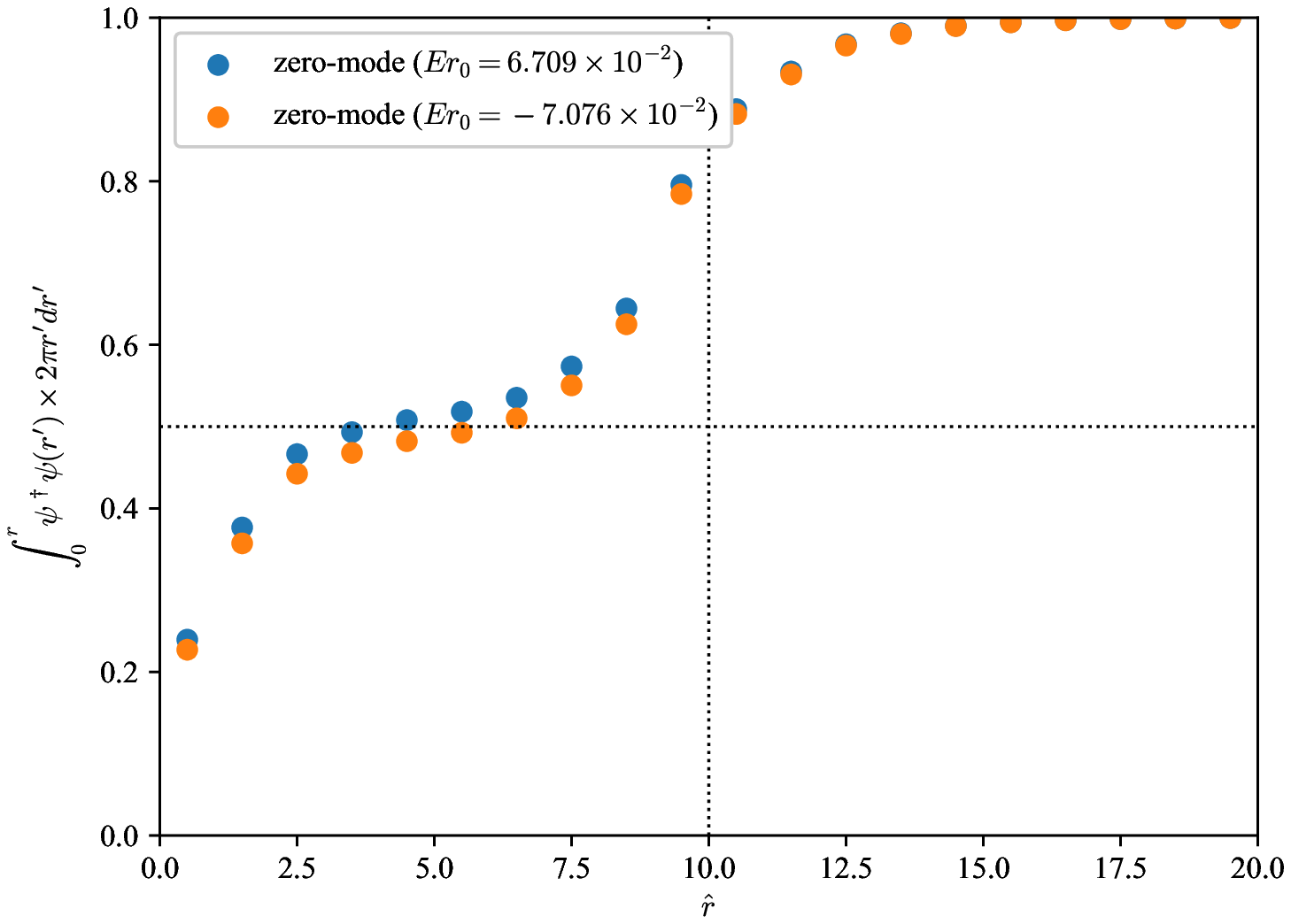}
  \end{minipage}
      \caption{Left panel shows the amplitude of two zero-modes in $r$ direction and right panel indicates the accumulation of that when $m= 14/L$, $\hat{r}_0=r_0/a = 10$, $\hat{r}_1=0.001$, $L = 40a$ and $\alpha=0.5$. The two zero-modes are localized at the wall and the origin by the tunneling effect. So their energy are slightly deviated from zero.}
    \label{fig:anmplitude in radius}
\end{figure}

Let us then set the Fermi energy to zero so that all negative states are occupied by the valence electrons. This is the so-called half-filled state. Then Fig.~\ref{fig:Anomaly inflow r1 =0.001*a} indicates that one conduction band localized at the $S^1$ domain-wall meets zero at $\alpha=0.5$ and one valence electron is pumped up to the zero energy at the center. By quantum tunneling effect, these two zero-modes mix and the energy levels are split, as Fig.~\ref{fig:anomaly inflow enlarge} shows. In Fig.~\ref{fig:anmplitude in radius}, we plot the distribution of $\psi^\dagger \psi(r)\times 2\pi r$ (left panel), and its integral up to $r$ (right panel). We can see that the $50$\% of the electron wave function is concentrated at the origin, where the point-like magnetic field is located.


We can interpret the above discussion as a microscopic description of the Witten effect. In the effective theory approach, the Witten effect is described by the Chern-Simons action, which induces an electric field to the monopole, becoming a dyon with charge $1/2$. In our lattice set up, we have shown that the point-like magnetic field (by the vortex) creates a domain-wall around it to capture  the electron wave function clinging to the defect. At $\alpha=1/2$ where the time-reversal symmetry is recovered, the electron is captured without any energy cost. Thus every vortex is charged.

It is important to note that another half of the electron wave function exists at the surface of the topological insulator, and the total electron charge is kept integral. It is also remarkable that the $U(1)$ gauge field  is singular at $x=-L/2,~L/2$ and $y=-L/2,~L/2$ due to the periodic boundary conditions on the lattice, but no localized modes appear near the singularity where the mass term is positive. 

Even if we remove the domain-wall, and set the mass negative everywhere, we can not find a zero mode localized at $x=-L/2,~L/2$ and $y=-L/2,~L/2$ since the flux at the singular edge is too weak to constrain a fermion. The singular edge breaks the $T$-symmetry so mod $2$ index can't classify this system.

\begin{figure}
    \centering
\includegraphics{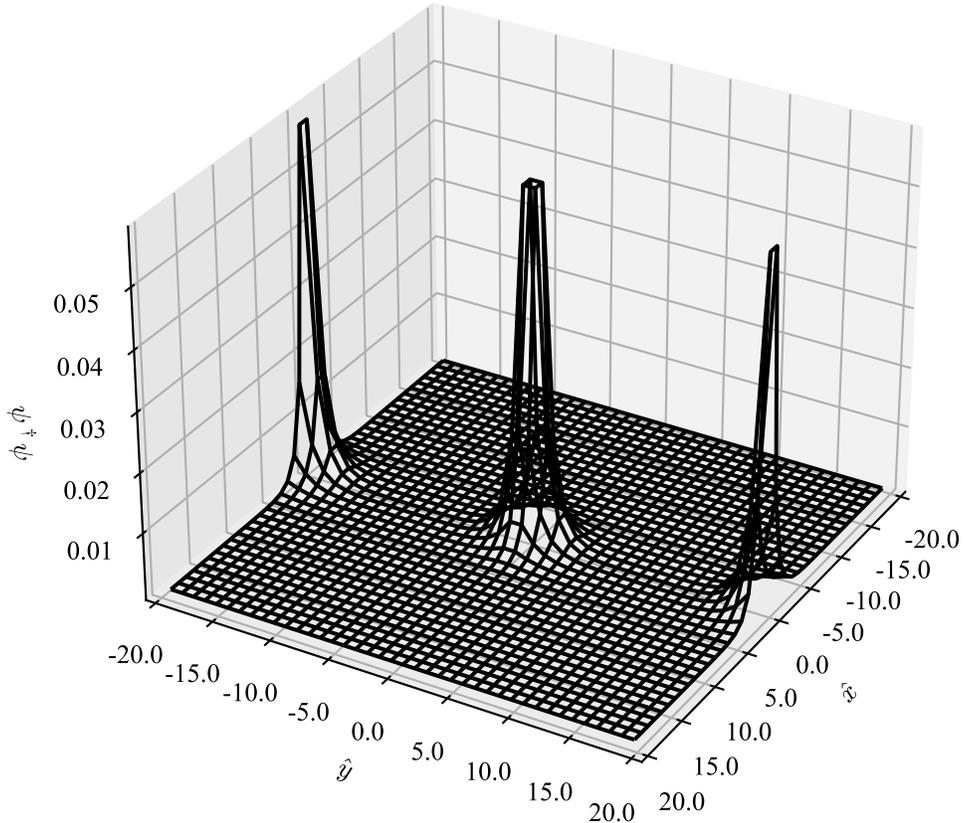}
    \caption{Two zero-modes appear at the vortex and anti-vortex.}
    \label{fig:noncontinuum flux}
\end{figure}

On the other hand, when we set a gauge field
\begin{align}
U_{1} (x,y)&=1 \\
U_{2} (x,y)&=\left\{ \begin{array}{cc}
    -1 & (x>0,~y=-\frac{a}{2}) \\
    +1 & (\text{otherwize})
\end{array} \right. .
\end{align}
We can maintain the $T$ symmetry. In this case, an anti-vortex appears at $(x,y)=(0, \frac{L}{2})$ in addition　to the original vortex. Each vortex binds one fermion with $E=0$ (see Fig. \ref{fig:noncontinuum flux}). Even number of zero-modes appear on the lattice space to cancel the anomaly.


%% file: main_S2_monopole.tex
\section{Anomaly of the $S^2$ domain-wall and Witten effect}
\label{sec:S2_Witten}

In this section, we consider a two-flavor $S^2$ domain-wall fermion on a three-dimensional square lattice. Because of the two-flavor structure having both of left- and right-handed chiralities, the edge modes represent a Dirac fermion field, rather than a Weyl fermion. The only possible anomaly they have is the standard axial $U(1)$ anomaly \cite{Adler1969Axial-Vector,Bell1969APCACpuzzle}, which gives a nontrivial AS index \cite{atiyah1963index,AtiyahSinger1986TheIndex1,AtiyahSegal1968TheIndex2,AtiyahSinger1968TheIndex3,AtiyahSinger1971TheIndex4} of the Dirac operator on $S^2$.

The AS index is, however, a cobordism invariant, which cannot be nonzero when the $S^2$ is a boundary of a three-dimensional bulk ball $B^3$. Therefore, when we force the domain-wall fermion to have a zero mode, we expect a dynamical domain-wall creation at the singularity of the gauge field to change the topology of the negative region of the fermion mass, and have another zero mode to cancel the axial $U(1)$ anomaly\footnote{The region where the mass is negative is topologically equivalent to $I\times S^2$, and there are two boundaries whose orientations are opposite to each other. Since the chirality is defined with the opposite sign for the two $S^2$, the $U (1)$ anomaly is canceled between them.
}.

The gauge field configuration which gives a nonzero index of the massless Dirac operator on $S^2$ is equivalent to putting a magnetic monopole \cite{Dirac:1931kp} inside the domain-wall. Therefore, if the field strength is strong enough to change the sign of the mass term from negative to positive near the monopole, creating another domain-wall, the appearance of the edge modes makes the monopole charged. This is exactly the same microscopic description as the previous section where a vortex is changed.

The Dirac monopole on the three-dimensional square lattice can be put by the $U(1)$ link variables assigned as
\begin{align}
    U_\mu (p)=\exp( i\int_{p+ a\hat{\mu}}^p A  )~(A=n\frac{1-\cos \theta}{2} d\phi),
\end{align}
where $n$ is a magnetic charge. In the same way as the two-dimensional lattice, we impose the periodic boundary conditions on the fermions, setting the link variables from connecting $x=N/2$ and $-N/2$  for all $y, z$, those connecting $y=N/2$ and $-N/2$ for all $x,z$ and those at $z=N/2$ for all $x,y$ to unity. The field strength produced around the boundary gives little effect on the edge-mode spectrum. Note in our formulation that the Dirac string located between $(0,0,0)$ and $(0,0,-L/2)$ is invisible as the plaquette becomes $\exp(2\pi n)=1$. This reflects the fact that the AB effect disappears when the magnetic flux is quantized.


With the above link configuration, let us consider the lattice Dirac operator
    \begin{align}
        H =\frac{1}{a}\gamma^5 \qty( \qty[\sum_{i=1}^3\gamma^i\frac{\nabla_i-\nabla^\dagger_i}{2} +\frac{1}{2}\nabla_i \nabla^\dagger_i ]+\epsilon_A am ), \label{eq:Hermitian Wilson Dirac op of S^2 in R^3}
    \end{align}
where we set $\gamma^i=\sigma_1 \otimes \sigma_i$ and $\gamma^5 =
\sigma_3 \otimes 1$.
The chirality operator is defined by
\begin{align}
    \gamma_{\text{normal}}= \frac{x}{r} \gamma^1 + \frac{y}{r} \gamma^2+\frac{z}{r} \gamma^3=\sigma_1 \otimes \frac{x^a \sigma_a}{r}.
\end{align}

\begin{figure}
    \centering
    \includegraphics{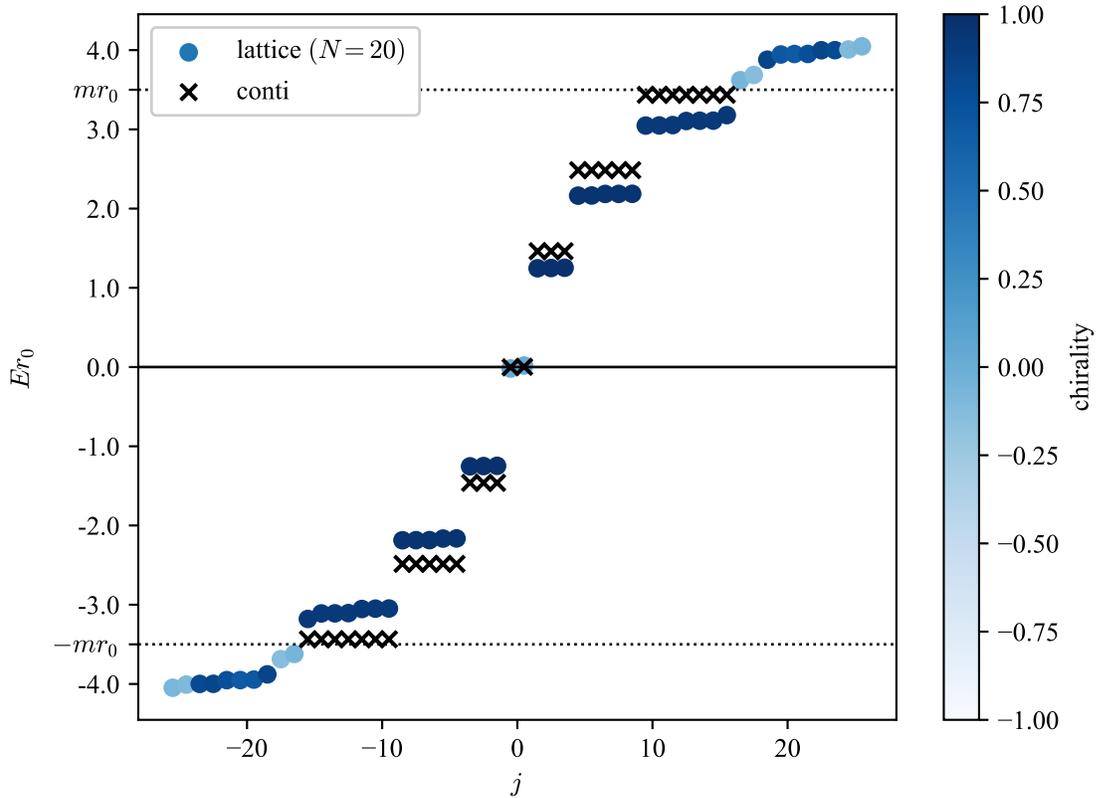}
    \caption{The spectrum of \eqref{eq:Hermitian Wilson Dirac op of S^2 in R^3} near $E=0$. We set $m=14/L$, $L=20a$, $r_0=\frac{L}{4}$ and $n=1$.}
    \label{fig:Eigenvalue S2flux gammanormal}
\end{figure}

In Fig.~\ref{fig:Eigenvalue S2flux gammanormal}, we plot the lattice data for the eigenvalues of $H$, where we set $m=14/L$, $L=20a$, $r_0=\frac{L}{4}$ and the monopole charge $n=1$. Similar to the $S^1$ domain-wall fermion case, the edge modes with $\gamma_\text{normal}=+1$ appear between $\pm m$.

Let us compare the result with the continuum theory in the large $m$ limit. 
The effective Dirac operator on $S^2$ on the edge modes is
\begin{align}\label{eq:Dirac op on S2}
     i\Slash{D}^{S^2}=i\qty(\sigma_1 \pdv{}{\theta} +\sigma_2
\frac{1}{\sin \theta} \qty( \pdv{}{\phi}+ \frac{i}{2}
-\frac{\cos\theta}{2 } \sigma_1 \sigma_2-in\frac{1-\cos \theta}{2 } )
) ,
\end{align}
where $\theta$ and $\phi$ represent polar angle and azimuthal angle, respectively \cite{Tamm1931monopole,WU1976365}. In the innermost parenthesis, the second and third terms are the induced gravitational $Spin^{(c)}$ connection. The fourth term is the $U(1)$ connection given by the monopole with magnetic charge $n$.

Since the angular momentum is conserved in the continuum theory, the eigenvalues labelled by its quantized numbers $j$ (of the angular momentum squared) and $j_3$ (in the $z$ direction), are obtained as \cite{Tamm1931monopole}
\begin{align}
    E_{j,j_3} r_0=\pm \sqrt{\qty(j+\frac{1}{2})^2 -\frac{n^2}{4}},~
j=\frac{\abs{n}-1}{2},~\frac{\abs{n}-1}{2}+1,\cdots,~j_3=-j,-j+1,\cdots +j,
\end{align}
where we have $2j+1$ degeneracies as usual. Note when $n=1$ and $j=0$, we have a zero mode. However, we can see two would-be zero modes in Fig.~ \ref{fig:Eigenvalue S2flux gammanormal}, which cannot be explained by the continuum Dirac operator on $S^2$.

\begin{figure}
    \centering
    \includegraphics{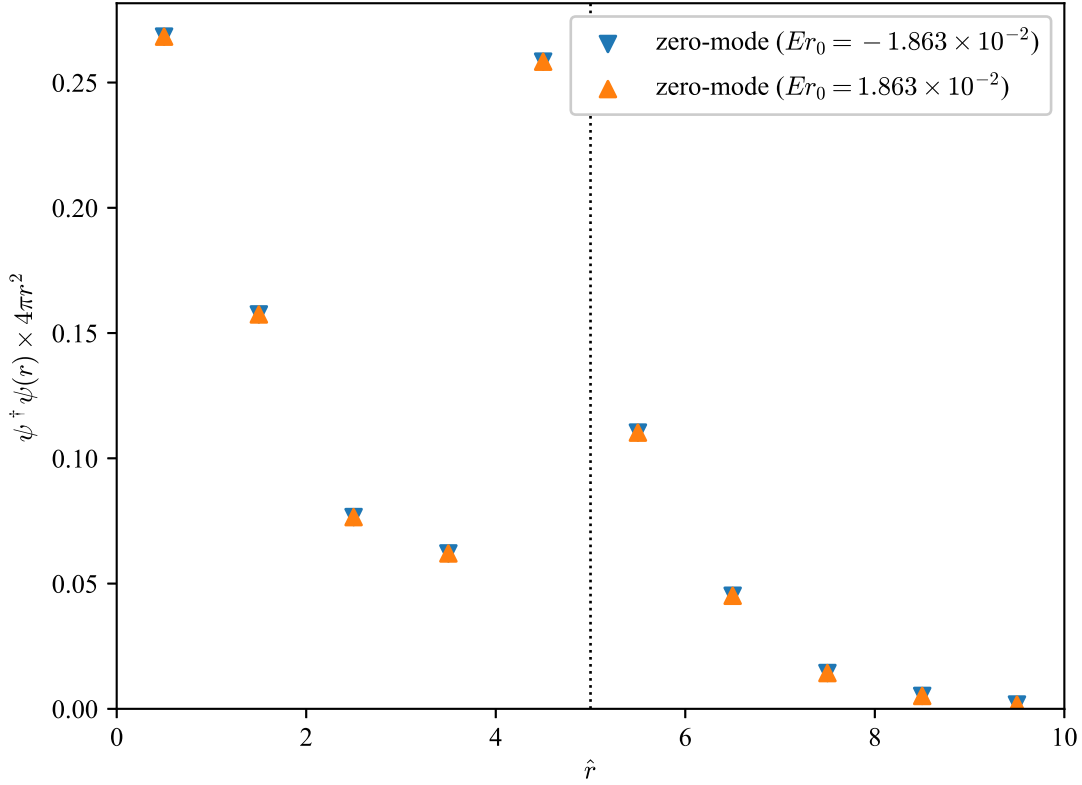}
    \caption{The amplitude of two zero-modes in $r$ direction when $m=14/L$, $L=20a$, $r_0 =L/4$ and $n=1$.}
    \label{fig:Amplitude S2 Edgemode and localizedmode_radius}
\end{figure}

As shown in Fig.~\ref{fig:Amplitude S2 Edgemode and localizedmode_radius},
these two near zero modes have peaks at $r=0$ and $r=r_0$.
which gives a strong evidence for the topology change of our
domain-wall fermion system
and  the two zero modes around $r=0$ and $r=r_0$
are mixed through the tunneling effect.
The charge the monopole dresses is thus, $\sim 1/2$.

\begin{figure}
    \centering
    \includegraphics{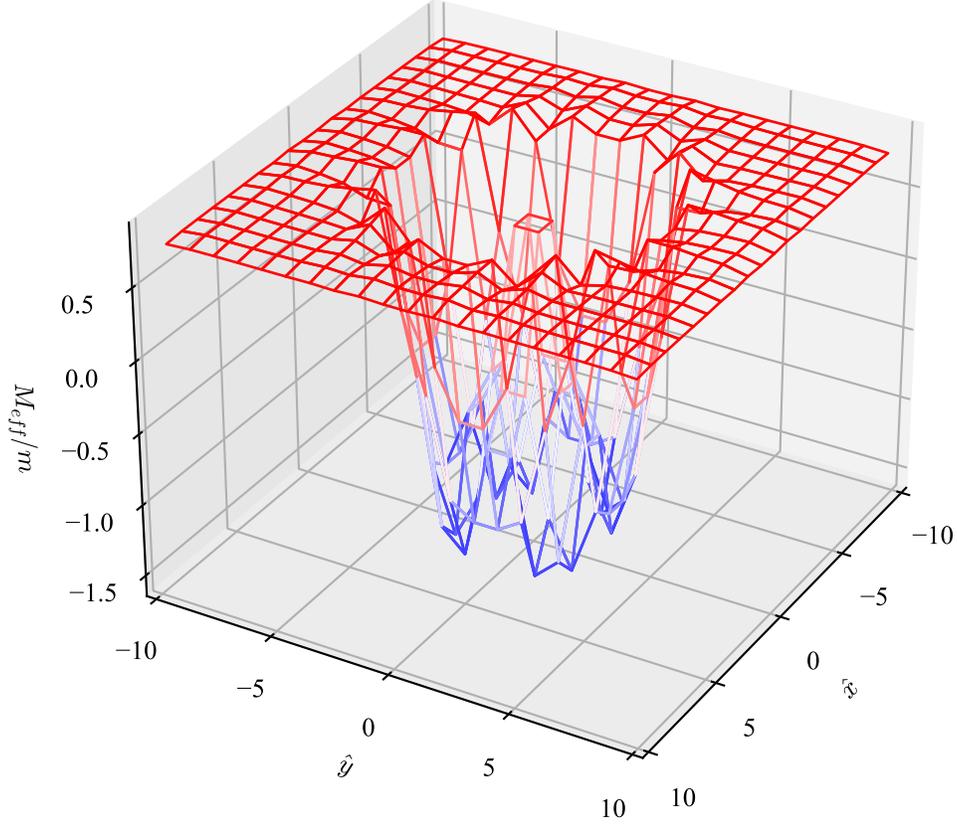}
    \caption{The ratio of the effective mass $M_{eff}$ and $m=\abs{M}$ when $m = 14/L$, $\hat{r}_0=r_0/a = 5$ and $L = 20 a$.}
    \label{fig:effective mass term S2}
\end{figure}

Let us also examine the effective mass shift due to the strong magnetic field. Fig.~\ref{fig:effective mass term S2} represents the position dependence of the effective mass,
\begin{align}
   M_{eff}=\frac{1}{ \psi(x)^\dagger \psi(x)} \psi(x)^\dagger
\qty(\epsilon m+\sum_{i=1}^3\frac{1}{2a}\nabla_i \nabla^\dagger_i  ) \psi(x)
\end{align}
which clearly indicates that another domain-wall is created near the origin, where the mass is shifted from negative to positive. Nothing special happens to the anti-monopole located in the normal insulator region.


Our numerical data above supports our  microscopic scenario for the Witten effect that the $U(1)$ gauge field from the monopole inside the topological insulator yields a mass shift, creating another domain-wall and the edge-localized mode is the origin of the electric charge. The electric charge of the dyon is 1/2 since another half is distributed at the surface of the topological insulator by the tunneling effect.